\documentclass[useAMS,usenatbib]{mn2e}
\usepackage{times,epsfig} 
\usepackage{natbib}
\usepackage{amssymb,amsmath}

\newcommand{\apj}{ApJ}
\newcommand{\apjl}{ApJ}
\newcommand{\apjs}{ApJS}
\newcommand{\aap}{A\&A}

\newcommand{\mnras}{MNRAS}
\newcommand{\nat}{Nature}

\newcommand{\prd}{Phys. Rev. D}

\title[$c-M$ relation for galaxy clusters]{
Hydrodynamical simulations of galaxy clusters in dark energy cosmologies: II. $c-M$ relation}

\author[C. De Boni et al.]
{\parbox[]{6.in} {C. De Boni$^1$\thanks{E-mail: cristiano.deboni@unibo.it}, S. Ettori$^{2,3}$, K. Dolag$^4$, L. Moscardini$^{1,2,3}$ \\
\footnotesize
$^1$ Dipartimento di Fisica e Astronomia, Universit\`a di Bologna, viale Berti Pichat 6/2,
  I-40127 Bologna, Italy \\
  $^2$ INAF, Osservatorio Astronomico di Bologna, via Ranzani 1,
  I-40127 Bologna, Italy  \\
$^3$ INFN, Sezione di Bologna, viale Berti Pichat 6/2,
  I-40127 Bologna, Italy  \\
$^4$ Max-Planck-Institut f\"ur Astrophysik, Karl-Scwarzschild Strasse 1, 
  D-85741 Garching bei M\"unchen, Germany \\
}}                                            
\date{}

\begin{document}

\maketitle

\begin{abstract}
We use hydrodynamical simulations of different dark energy cosmologies to investigate the concentration-mass ($c-M$) relation in galaxy clusters. In particular, we consider a reference $\Lambda$CDM model, two quintessence models with inverse power-law potentials (RP and SUGRA), and two extended quintessence models, one with positive and one with negative coupling with gravity (EQp and EQn respectively). All the models are normalized in order to match CMB data from WMAP3. We fit both the dark matter only and the total mass profile with a NFW profile, and recover the concentration of each halo from the fit using different definition. We consider both the complete catalog of clusters and groups and subsamples of objects at different level of relaxation. We find that the definition itself of the concentration can lead to differences up to $20\%$ in its value and that these differences are smaller when more relaxed objects are considered. The $c-M$ relation of our reference $\Lambda$CDM model is in good agreement with the results in literature, and relaxed objects have a higher normalization and a shallower slope with respect to the complete sample. The inclusion of baryon physics is found to influence more high-mass systems than low-mass ones, due to a higher concentration of baryons in the inner regions of massive halos. For the different dark energy models, we find that for $\Lambda$CDM, RP and SUGRA the normalization of the $c-M$ relation is linked to the growth factor, with models having a higher value of $\sigma_{8} D_{+}$ having also a higher normalization. This simple scheme is no longer valid for EQp and EQn because in these models a time dependent effective gravitational interaction, whose redshift evolution depends on the sign of the coupling, is present. This leads to a decrease (increase) of the expected normalization in the EQp (EQn) model. This result shows a direct manifestation of the coupling between gravity and the quintessence scalar field characterizing EQ models that can be in principle investigated through the analysis of the $c-M$ relation.
\end{abstract}


\section{Introduction}

The internal properties of dark matter halos are known to reflect their formation history and thus the evolution of the background cosmology. \cite*{1996ApJ...462..563N} (hereafter NFW) found that the dark matter profile of a halo can be characterized by a scale radius, which is linked to the virial radius through the concentration of the object. The concentration of a dark matter halo is related to the mean density of the universe at the halo formation time. Moreover, the presence of baryons influences these internal properties because they tend to concentrate in the inner regions, increasing thus the concentration of the halo. 

Because of the hierarchical nature of structure formation and the fact that collapsed objects retain information on the background average matter density at the time of their formation \citep{1996ApJ...462..563N}, concentration and mass of a dark matter halo are related. Since low-mass objects form earlier than high-mass ones, and since in the past the background average matter density was higher, low-mass halos are expected to have a higher concentration compared to high-mass ones. For a given mass, the evolution of the concentration with redshift is still matter of debate \citep[see {\it{e.g.}}][]{2001MNRAS.321..559B,2001ApJ...554..114E,2007MNRAS.381.1450N,2008MNRAS.390L..64D,2002ApJ...574..538J,2006MNRAS.367.1781A,2011arXiv1104.5130P,2012MNRAS.419.2821C}. These expectations have been confirmed by the results of $N$-body numerical simulations which find, at $z=0$, a concentration-mass relation $c(M) \propto M^{\alpha}$, with $\alpha \sim -0.1$ \citep{2004A&A...416..853D,2008MNRAS.387..536G,2010ApJ...712L.179Z}, with a log-normal scatter ranging from $0.15$ for relaxed systems to $0.30$ for disturbed ones \citep{2000ApJ...535...30J}. This is true for idealized, dark matter only halos, but in real objects the impact of baryon physics is strong, in particular in the inner regions. Indeed, even if the ICM is described by a cored $\beta$-model profile, in the central region it tends to cool and to form stars, which accumulate in the core of the cluster, thus increasing the total concentration. Moreover, it is supposed that this baryon contraction does also affect the dark matter component, through an adiabatic contraction effect, leading thus to an increase of the dark matter concentration. The adiabatic contraction model has been described in \cite{1986ApJ...301...27B} and \cite{1987ApJ...318...15R}, but, for an updated treatment, see also \cite{2012MNRAS.tmp.3239F}. 

$N$-body simulations have been carried out by several authors in order to study the $c-M$ relation in dark matter halos with sizes of galaxy groups and clusters. \cite{2004A&A...416..853D} performed simulations with different cosmological models in order to verify the effects of dark energy dynamics. For $\Lambda$CDM they found that the dependence of concentration on mass and redshift can be fitted by a relation of the form $(1+z) c(M) = M^{\alpha}$, with $\alpha \approx -0.1$. For dark energy models, they found that the halo concentration depends on the dark energy equation of state through the linear growth factor at the cluster formation redshift, $D_{+}(z_{coll})$. \cite{2007MNRAS.381.1450N} also noted that non-relaxed objects have a lower concentration and a higher scatter with respect to relaxed ones. \cite{2008MNRAS.391.1940M} made a comparison between concentrations in the WMAP1, WMAP3 and WMAP5 cosmologies in order to study the effects of different cosmological parameters (in particular $\sigma_{8}$) on the $c-M$ relation. \cite{2008MNRAS.390L..64D} found that the concentrations of their halos is lower than the one inferred from X-ray observations and addressed that fact to the effect of baryon physics that was missing in their simulations. Actually, despite an initial compatibility of the results in \cite{2005A&A...429..791P} and \cite{2005A&A...435....1P} with the predictions of \cite{2004A&A...416..853D}, considering the most recent data sets there is still poor agreement between the observed $c-M$ relation and the predicted one, with the first having, in general, a steeper slope and a higher normalization compared to the latter \citep{2007MNRAS.379..190C,2007ApJ...664..123B,2007MNRAS.379..209S,2010A&A...524A..68E,2010MNRAS.408.2442W,2012MNRAS.420.3213O,2012arXiv1201.1616C}. Numerical simulations to study the impact of baryon physics on the structure of dark matter halos, and to try to reconcile the discrepancy between the expectations from dark matter only simulations and the concentrations inferred from observations, were carried out in \cite{2010MNRAS.405.2161D}. However, they showed that even including baryon physics in the simulations they cannot reproduce both observed concentrations and stellar fraction in galaxy groups and clusters. In general, including baryons they found lower concentrations with respect to the dark matter only case. The triaxiality of the matter distribution and the presence of substructures within the host halo virial radius could bias the estimates of the mass and concentration obtained through lensing techniques at the level of few per cent, but cannot reconcile the observed and simulated relations \citep[see][]{2012arXiv1205.2375G}.

Since the concentration of a halo is linked to the background density of the universe at the time it collapsed, and since different dark energy models predict different evolutions of the cosmological background, it is interesting to investigate the impact of dark energy on the $c-M$ relation. Moreover, since some dark energy models can also affect the linear and non linear evolution of the density fluctuations, leaving some imprints in collapsed structure, one can think about using the $c-M$ relation as a cosmological probe, orthogonal to others that are commonly used. $N$-body cosmological simulations of extended quintessence models, including the effects on the $c-M$ relation, were presented in \cite{2011ApJ...728..109L}. \cite{2009PhRvD..80d3530B} performed a theoretical study on the spherical collapse model beyond $\Lambda$CDM and found that the observed $c-M$ relation could indicate the possibility for a clustered dark energy.  

In this paper we address this issue by studying the $c-M$ relation in galaxy clusters extracted from dark matter only and hydrodynamical simulations of five dark energy models introduced in \cite{2011MNRAS.415.2758D} (hereafter Paper I). These simulations allow us to evaluate at the same times the effects of baryon physics and dynamics of dark energy on the internal properties of galaxy clusters.

The paper is organized as follows. Dark energy models and numerical simulations are presented in Sect. \ref{models}. After introducing different methods to define the concentration in Sect. \ref{fitting}, along with a non-parametric definition in Sect. \ref{nomodel}, we compare our results for the $\Lambda$CDM model with other results from literature in Sect. \ref{comparison}. For the different dark energy cosmologies, we analyse the dark matter only runs in Sect. \ref{dark_matter} and the hydrodynamical runs in Sect. \ref{total}. We discuss the results in Sect. \ref{other} and draw our conclusions in Sect. \ref{c-M_relation_summary}.

\section{The cosmological models} \label{models}

We consider the same cosmological models discussed in Paper I. Here we recall only the main features of the different models, and refer to Paper I for more details.

As a reference model we use the concordance $\Lambda$CDM model, adapted to the WMAP3 values \citep{2007ApJS..170..377S}, with the following cosmological parameters:

\begin{itemize}
\item matter density: $\Omega_{0m}=0.268$
\item dark energy density: $\Omega_{0\Lambda}=0.732$
\item baryon density: $\Omega_{0b}=0.044$
\item Hubble parameter: $h=0.704$
\item power spectrum normalization: $\sigma_{8}=0.776$
\item spectral index: $n_{s}=0.947$
\end{itemize}

This model is characterized by the presence of a dark energy component given by a cosmological constant $\Lambda$, with a constant $w_{\Lambda}=-1$. 

The second case is a model with dynamical dark energy, given by a quintessence scalar field $\phi$ with an equation of state $w=w(a)$ \citep{1988NuPhB.302..645W,1988PhRvD..37.3406R}. As in Paper I, as potentials for minimally coupled quintessence models, we consider an inverse power-law potential

\begin{equation}
\label{rp_potential}
V(\phi)=\frac{M^{4+\alpha}}{\phi^{\alpha}} \ ,
\end{equation}

\noindent the so called RP potential \citep{1988PhRvD..37.3406R}, as well as its generalization suggested by supergravity arguments \citep{1999PhLB..468...40B}, known as SUGRA potential, given by

\begin{equation}
\label{sugra_potential}
V(\phi)=\frac{M^{4+\alpha}}{\phi^{\alpha}}\exp(4\pi G \phi^2) \ ,
\end{equation}

\noindent where in both cases $M$ and $\alpha \ge 0$ are free parameters (see Table \ref{tab} for details).

The third possibility we consider is the case in which $\phi$ interacts non minimally with gravity \citep{1988NuPhB.302..645W,2000PhRvL..85.2236B}. In particular we refer to the extended quintessence (EQ) models described in \cite{2000PhRvD..61b3507P}, \cite{2005JCAP...12..003P} and \cite{2008PhRvD..77j3003P}. The parameter $\xi$ represents the "strength" of the coupling (see Table \ref{tab} for details). In particular we consider here a model with positive coupling $\xi > 0$ (EQp) and one with negative $\xi < 0$ (EQn). For an extensive linear treatment of EQ models we refer to
\citet{2008PhRvD..77j3003P}. Here we only recall for convenience that
EQ models behave like minimally coupled quintessence theories in
which, however, a time dependent effective gravitational interaction
is present. In particular, in the Newtonian limit, the gravitational
parameter is redefined as

\begin{equation} 
\label{EQ_Gtilde_def} 
\tilde{G} = \frac{2 [ F + 2(\partial F/\partial \phi)^2]}{[ 2 F+3(\partial F/\partial \phi)^2]} \frac{1}{8 \pi F} \ .
\end{equation} 

\noindent Here the coupling $F(\phi)$ is chosen to be

\begin{equation}
\label{non-minimal coupling}
F(\phi)=\frac{1}{\kappa}+\xi (\phi^2 - {\phi_{0}^2}) \ ,
\end{equation}

\noindent with $\kappa=8\pi G_\ast$, where $G_{\ast}$ represents the ``bare'' gravitational constant \citep{2001PhRvD..63f3504E}.

\begin{figure}
\hbox{
 \epsfig{figure=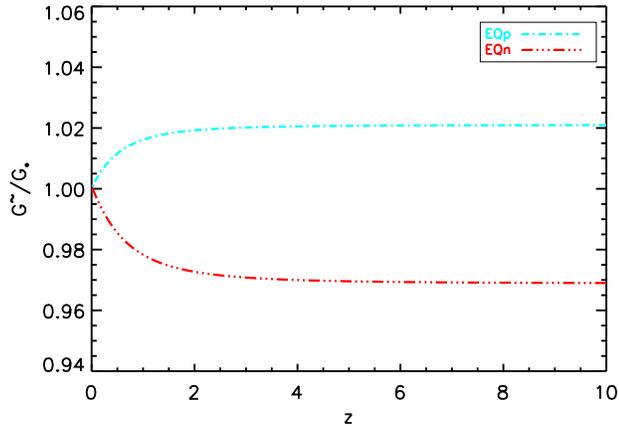,width=0.50\textwidth}
}
\caption{Correction to the gravity constant for the two extended quintessence models, EQp (cyan) and EQn (red), as expressed in equation (\ref{dG}).}
\label{dG_z}
\end{figure}

\noindent For small values of the coupling, that is to say $\xi \ll 1$, the latter expression becomes

\begin{equation}  
\frac{\tilde{G}}{\ G_\ast} \sim 1 - 8 \pi G_\ast \xi (\phi^2 - {\phi_{0}^2}) \ ,
\label{dG}
\end{equation}

\noindent which manifestly depends on the sign of the coupling $\xi$. We note that, since the derivative of the RP potential in equation (\ref{rp_potential}) with respect to $\phi$ is $\partial V(\phi)/ \partial \phi < 0$, we have $\phi^2 < {\phi_{0}^2}$. This leads to the behaviour of ${\tilde{G}}/{G_\ast}$ shown in Fig. \ref{dG_z}. Note that the corrections are only within the percent level.

For the $\Lambda$CDM model, the linear density contrast of a top-hat spherical collapse model, $\delta_{c}$, converges to the Einstein-de Sitter value $1.686$ at high redshift, but it is lower than the Einstein-de Sitter value at low redshift due to the presence of the cosmological constant which starts to dominate. For minimally coupled and extended quintessence models, the value of $\delta_{c}$ at low redshift is lower than the $\Lambda$CDM one, depending on the model. In any case, at $z=0$ these differences are very small, well below $1\%$. For an extensive treatment of EQ models see Pace et al. (in prep.). \\

We trimmed the parameters of the four dynamical dark energy models in such a way that $w_0=w(0)\approx-0.9$ is the highest value still consistent with present observational constraints in order to amplify the effects of dark energy. The parameters $\Omega_{0m}$, $\Omega_{0\Lambda}$, $\Omega_{0b}$, $h$, and $n_{s}$ are the same for all models, but since we normalize the power spectrum to CMB data from WMAP3, this leads to different values of $\sigma_{8}$ for the different cosmologies: 

\begin{equation}
\sigma_{8,\mathrm{DE}} = \sigma_{8,\mathrm{\Lambda CDM}} \frac{D_{+,\mathrm{\Lambda CDM}}(z_\mathrm{CMB})}{D_{+,\mathrm{DE}}(z_\mathrm{CMB})} \ ,
\label{sigmaDE}
\end{equation}

\noindent assuming $z_\mathrm{CMB}=1089$, where the growth factor $D_{+}$ is normalized to unity today. In principle, in EQ models there could be a scale-dependent linear growth rate which is neglected in equation (\ref{sigmaDE}). Given the values of the coupling, we do expect this dependence to be small. Table \ref{tab} lists the parameters chosen for the different cosmological models.

\begin{table}
\caption{Parameters for the different cosmological models: $\alpha$ is the exponent of the inverse power-law potential; $\xi$ is the coupling in the extended quintessence models; $w_0$ is the present value of the equation of state parameter for dark energy; $\sigma_{8}$ is the normalization of the power spectrum.}
\begin{tabular}{|l|c|c|c|c|}
\hline
Model & $\alpha$ & $\xi$ & $w_0$ & $\sigma_{8}$ \\ 
\hline
$\Lambda$CDM & --- & ---  & $-1.0$ & $0.776$ \\
RP & $0.347$ & ---   & $-0.9$ & $0.746$ \\
SUGRA & $2.259$ & ---  & $-0.9$  & $0.686$ \\
EQp & $0.229$ & $+0.085$ & $-0.9$  & $0.748$ \\ 
EQn  & $0.435$ & $-0.072$ & $-0.9$  & $0.729$ \\
\hline
\end{tabular}
\label{tab}
\end{table}

In order to study the formation and evolution of large scale structures in
these different cosmological scenarios we use $N$-body and $N$-body + hydrodynamical
simulations performed with the {\small{GADGET-3}} code
\citep{2001ApJ...549..681S,2005MNRAS.364.1105S}, 
which makes use of the entropy-conserving formulation of SPH
\citep{2002MNRAS.333..649S}. The hydrodynamical simulations include radiative cooling,
heating by a uniform redshift--dependent UV background
\citep{1996ApJ...461...20H}, and a treatment of star formation and
feedback processes. For the dark matter only simulations, we simulated a cosmological box of size $(300 \ {\rm{Mpc}} \ h^{-1})^{3}$, resolved with $(768)^{3}$ dark matter particles with a mass of 
$m_{dm} \approx 4.4 \times 10^{9}  \ {\rm{M_{\odot}}} \ h^{-1}$. For the hydrodynamical simulations, we considered a cosmological box of size $(300 \ {\rm{Mpc}} \ h^{-1})^{3}$, resolved with $(768)^{3}$ dark matter particles with a mass of 
$m_{dm} \approx 3.7 \times 10^{9}  \ {\rm{M_{\odot}}} \ h^{-1}$ and the same amount of gas particles, having a mass of $m_{gas} \approx 7.3 \times 10^{8}  \ {\rm{M_{\odot}}} \ h^{-1}$.

\begin{figure}
\hbox{
 \epsfig{figure=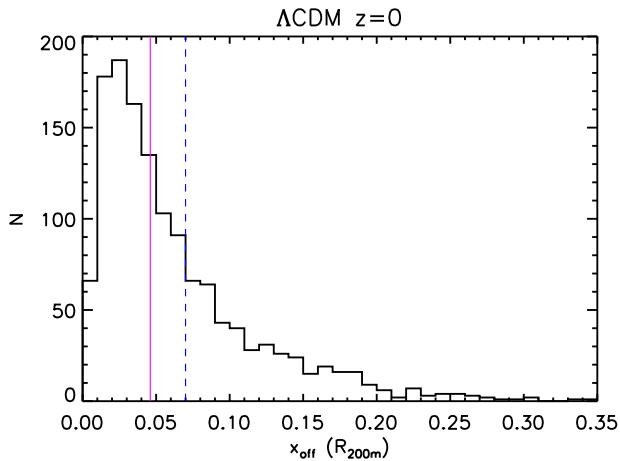,width=0.50\textwidth}
}
\caption{The values of $x_{off}$ in units of $R_{200m}$ for the objects in $\Lambda$CDM at $z=0$. The vertical dashed blue line corresponds to the value defining relaxed objects, $x_{off}=0.07R_{200m}$. The vertical pink line marks the median value, $x_{off}=0.046R_{200m}$.}
\label{xoff_histogram}
\end{figure}

Using the outputs of simulations, we extract galaxy clusters from the cosmological boxes, using the spherical overdensity criterion to define the collapsed structures. We take as halo centre the position of the most bound particle. Around this particle, we construct spherical shells of matter and stop when the total ({\it{i.e.}} dark matter plus gas plus stars) overdensity drops below $200$ times the {\it mean} (as opposed to {\it critical}) background density defined by $\Omega_{m}\rho_{0c}$; the radius so defined is denoted with $R_{200m}$ and the mass enclosed in it as $M_{200m}$. We consider all the halos having $M_{200m}> 10^{14} \ {\rm{M_{\odot}}} \ h^{-1}$. In addition, we selected subsamples of the $200$ objects with $M_{200m}$ closest to $7 \times 10^{13} \ {\rm{M_{\odot}}} \ h^{-1}$, $5 \times 10^{13} \ {\rm{M_{\odot}}} \ h^{-1}$, $3 \times 10^{13} \ {\rm{M_{\odot}}} \ h^{-1}$, and $10^{13} \ {\rm{M_{\odot}}} \ h^{-1}$. Starting from the centres of the halos, we construct radial profiles by binning the particles in radial bins. We select and study objects at three different redshifts: $z=0$, $z=0.5$, and $z=1$. For the following analysis, we also calculate for each cluster selected in this way the radius at which the overdensity drops below $200$ ($500$) times the {\it critical} background density and denote it as $R_{200}$ ($R_{500}$). The corresponding mass is indicated as $M_{200}$ ($M_{500}$). It is useful to define a quantitative criterion to decide whether a cluster can be considered relaxed or not because, in general, relaxed clusters have more spherical shapes, better defined centres and thus are more representative of the self-similar behaviour of the dark matter halos. We use a simple criterion similar to the one introduced in \cite{2007MNRAS.381.1450N}: first we define $x_{off}$ as the distance between the centre of the halo (given by the most bound particle) and the barycentre of the region included in $R_{200m}$; then we define as relaxed the halos for which $x_{off}< 0.07R_{200m}$. We plot the distribution of $x_{off}$ for the objects in $\Lambda$CDM at $z=0$ in Fig. \ref{xoff_histogram}.

\section{Fitting $\lowercase{c}-M$ relation} \label{fitting}

In this paper, we analyse both the dark matter only and the hydrodynamical runs of the simulation set introduced in Paper I. We calibrate the fitting procedure on $\Lambda$CDM because we use it as a reference model, and most references in the literature are based on this model.

For the concordance $\Lambda$CDM model, for each cluster at $z=0$ in the dark matter only run, we perform a logarithmic fit, using Poissonian errors $(\ln 10 \times \sqrt{n_{dm}})^{-1}$ (where $n_{dm}$ is the number of dark matter particles in each radial bin, of the order of $10-10^3$ depending on the mass of the object), of the three-dimensional dark matter profile $\rho_{dm}(r)$ in the region [$0.1-1$]$R_{200}$ (where the value of $R_{200}$ is taken directly from the true mass profile) with a NFW profile \citep{1996ApJ...462..563N}

\begin{equation}
\frac{\rho_{dm}(r)}{\rho_c}=\frac{\delta}{(r/r_{s})(1+r/r_{s})^{2}} \ ,
\label{NFW_c-M}
\end{equation}

\noindent where $\rho_{c}$ is the critical density, $r_{s}$ is the scale radius and $\delta$ is a characteristic density contrast. Then, instead of defining $c_{200} \equiv R_{200}/r_{s}$, we directly find the concentration parameter $c_{200}$ from the normalization of the NFW profile

\begin{equation}
\delta=\frac{200}{3}\frac{c_{200}^3}{\left[\ln (1+ c_{200}) - \frac{c_{200}}{1+c_{200}}\right]} \ .
\label{c_fit}
\end{equation}

\begin{figure}
\hbox{
 \epsfig{figure=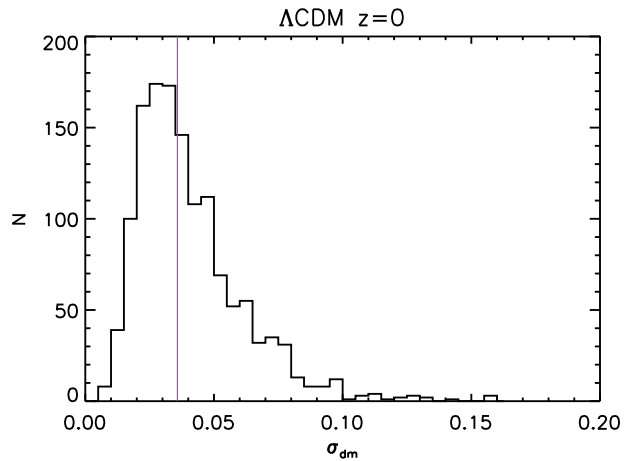,width=0.50\textwidth}
}
\caption{The values of $\sigma_{rms}$ for the objects in $\Lambda$CDM at $z=0$. The vertical pink line marks the median value, $\sigma_{rms}=0.0358$.}
\label{sigma_histogram}
\end{figure}

\noindent We require the central density parameter $\delta$ to be greater than $100$ and the scale radius $r_{s}$ to be within [$0.1-1$]$R_{200}$. We exclude the inner regions from the fit because we are limited in resolution inside a given radius. We indicate the dark matter concentration found in this way as $c_{200dm}$. We define the rms deviation $\sigma_{rms}$ as

\begin{equation}
\sigma^2_{rms} = \frac{1}{N_{bins}} \sum_{i=1}^{N_{bins}} [{\rm{log_{10}}} \rho_{i} - {\rm{log_{10}}} \rho_{NFW}]^2 \ ,
\label{sigma}
\end{equation}

\noindent where $N_{bins}$ is the number of radial bins over which the fit is performed. We plot the distribution of $\sigma_{rms}$ for the objects in $\Lambda$CDM at $z=0$ in Fig. \ref{sigma_histogram}.

In addition to the complete and relaxed samples discussed in Sect. \ref{models}, we create a sample of ``super-relaxed'' objects defined as the ones having both $x_{off}$ and $\sigma_{rms}$ lower than the median value extracted from the complete sample, namely $0.046$ and $0.0358$, respectively. We do this to check if strict restrictions on both the dynamical state and the shape of the profile of the objects can reduce the intrinsic scatter in the values of concentration. We stress that the definition of super-relaxed objects is dependent on the way we fit the profile. In the end, the complete sample, the relaxed sample and the super-relaxed sample at $z=0$ are constituted by 1357, 923, and 411 objects, respectively.

\noindent We bin the objects in the complete sample in groups of 200, so that we have bins around $10^{13} \ {\rm{M_{\odot}}} \ h^{-1}$, $3 \times 10^{13} \ {\rm{M_{\odot}}} \ h^{-1}$, $5 \times 10^{13} \ {\rm{M_{\odot}}} \ h^{-1}$, and $7 \times 10^{13} \ {\rm{M_{\odot}}} \ h^{-1}$. For halos more massive than $10^{14} \ {\rm{M_{\odot}}} \ h^{-1}$, we bin the objects starting from the low-mass ones, so that the most massive bin can contain less than 200 objects. The analysis for the relaxed and super-relaxed samples is done selecting the relaxed and super-relaxed objects inside each bin. Once we have $c_{200dm}$ for each object in each mass bin, since the distribution of $c_{200dm}$ is log-normal inside each bin, we evaluate the mean $M_{200}$ and the mean and rms deviation of ${\rm{log_{10}}}  c_{200dm}$ in each bin, for all the three samples. In the following of the paper, when we indicate the value of $c_{200dm}$ in a mass bin, we refer to $10^{\langle{{\rm{log_{10}}} c_{200dm}}\rangle}$.

\begin{figure}
\hbox{
 \epsfig{figure=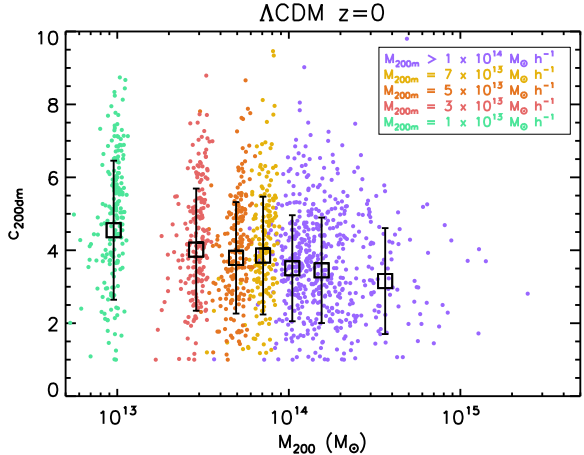,width=0.50\textwidth}
}
\caption{The values of $c_{200dm}$ for the complete sample of the $\Lambda$CDM model at $z=0$. For each object, we plot $M_{200}$ and $c_{200dm}$. Different colours indicate different $M_{200m}$ ranges. The black squares indicate $c_{200dm}$ in each mass bin (see text for details). For each mass bin, we plot the mean $M_{200}$ and $c_{200dm}$ with rms deviation.}
\label{c_dm-M}
\end{figure}

\noindent In Fig. \ref{c_dm-M} we plot $c_{200dm}$ for each object in the complete sample. We see that there is a large intrinsic dispersion in the values of the concentration inside each group of objects, which is marked with a different colour. For objects more massive than $10^{14} \ {\rm{M_{\odot}}} \ h^{-1}$, the maximum value of $c_{200dm}$ clearly decreases with $M_{200}$. Only when we bin the objects and plot the mean value of $c_{200dm}$ in each mass bin we can see a general trend with the concentration decreasing with increasing mass, even if the rms deviations are quite large.

\begin{figure*}
\hbox{
 \epsfig{figure=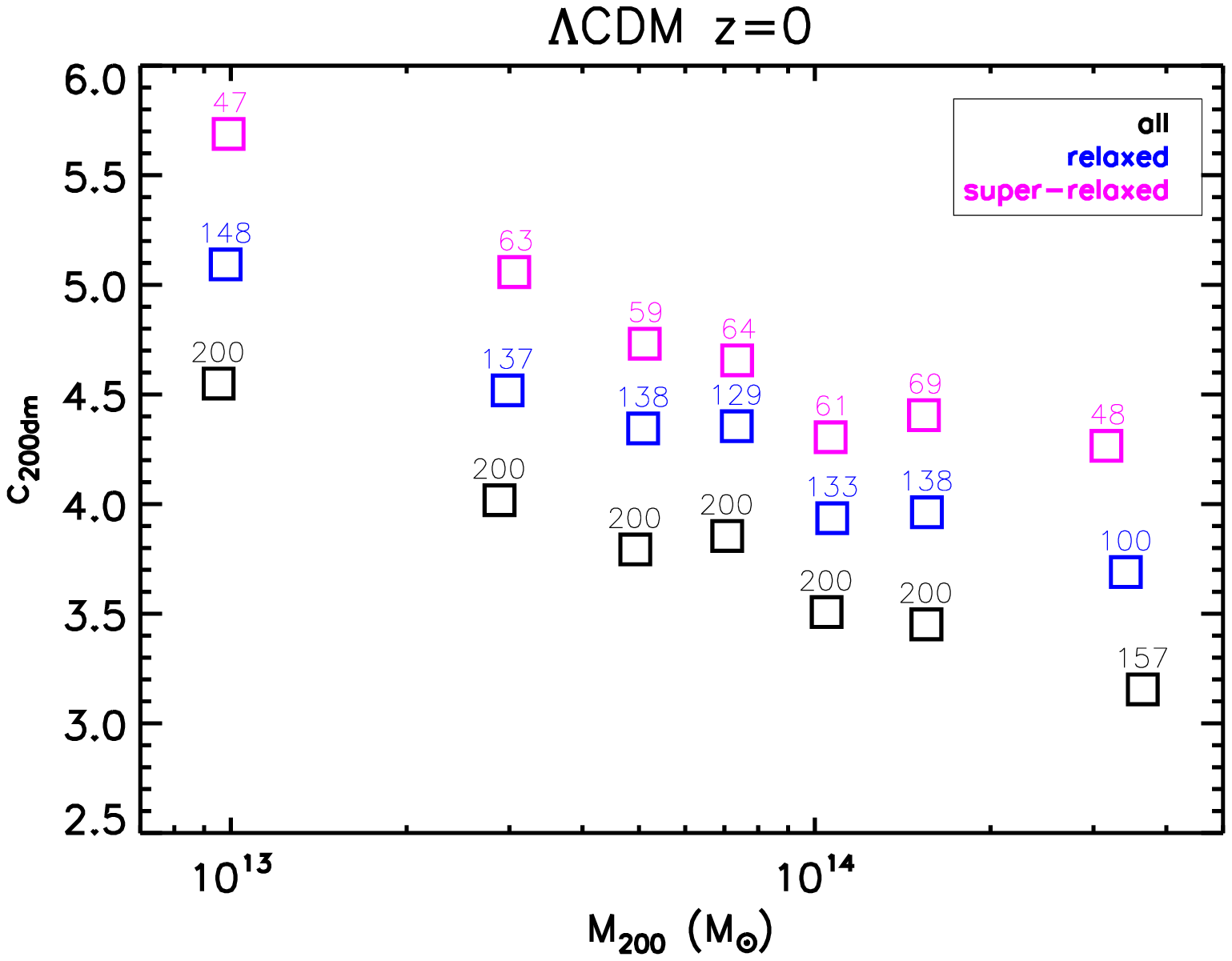,width=0.50\textwidth}
 \epsfig{figure=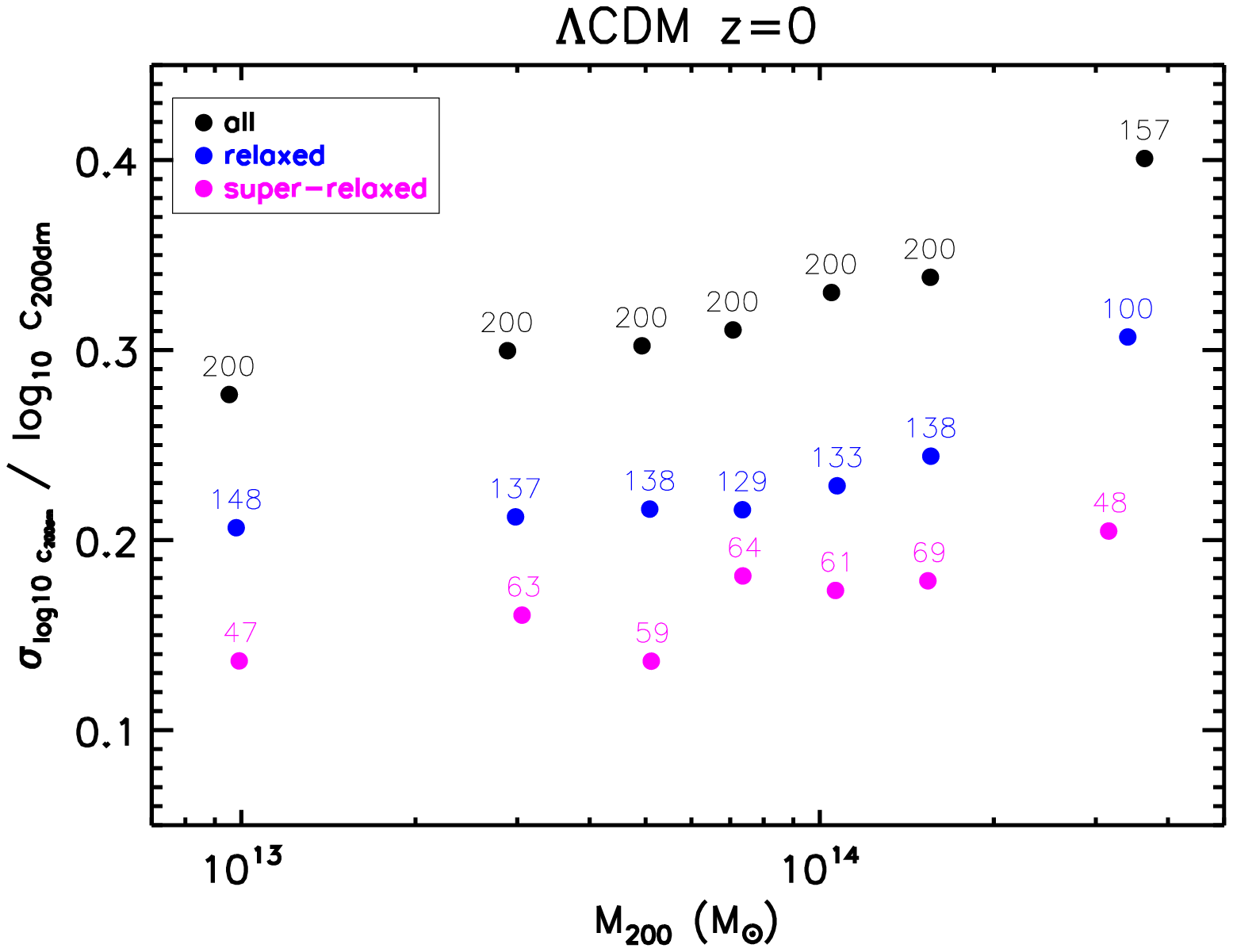,width=0.50\textwidth}
}
\caption{(Left panel) The values of $c_{200dm}$ for the complete (black), relaxed (blue) and super-relaxed (pink) samples of the $\Lambda$CDM model at $z=0$. For each mass bin, we plot the mean $M_{200}$, $c_{200dm}$ and the number of objects in the bin. (Right panel) Relative error in ${\rm{log_{10}}}  c_{200dm}$ for the complete (black), relaxed (blue) and super-relaxed (pink) samples of the $\Lambda$CDM model at $z=0$. For each mass bin, we plot the mean $M_{200}$, the relative error $\sigma_{{\rm{log_{10}}}  c_{200dm}} /{\rm{log_{10}}}  c_{200dm}$ and the number of objects in the bin.}
\label{super-relaxed_sdev_comparison}
\end{figure*}

\noindent In Fig. \ref{super-relaxed_sdev_comparison} we plot the values of$c_{200dm}$ in each mass bin for the complete, relaxed and super-relaxed samples, along with the number of objects in each bin. We note that, in all bins, more than $50\%$ of the halos are relaxed and around $30\%$ are super-relaxed. In each bin, the value of $c_{200dm}$ for the relaxed sample is higher than the one for the complete sample, and the value for the super-relaxed sample is even higher. In all three samples, $c_{200dm}$ is decreasing with increasing mass. In Fig. \ref{super-relaxed_sdev_comparison} we also show the scatter in ${\rm{log_{10}}}  c_{200dm}$ in each bin for the complete, relaxed and super-relaxed samples. For the complete sample, the scatter ranges from $30\%$ to $40\%$, with a positive trend with mass. It means that inside each mass bin there are objects with quite different concentrations, in particular in the high-mass tail. For the relaxed sample, the scatter reduces to $20\%$ up to $30\%$, meaning that part of the scatter in the complete sample is due to objects in a particular dynamical state. If we move to the super-relaxed sample, we notice that the scatter stays between $15\%$ and $20\%$. On the one hand, this means that putting strong constraints on both the dynamical state and the shape of the profile of the halos allows us to halve the intrinsic scatter; on the other hand, even when considering the most relaxed and smooth objects in our sample we cannot reduce the intrinsic scatter below $15\%$.

\begin{figure*}
\hbox{
 \epsfig{figure=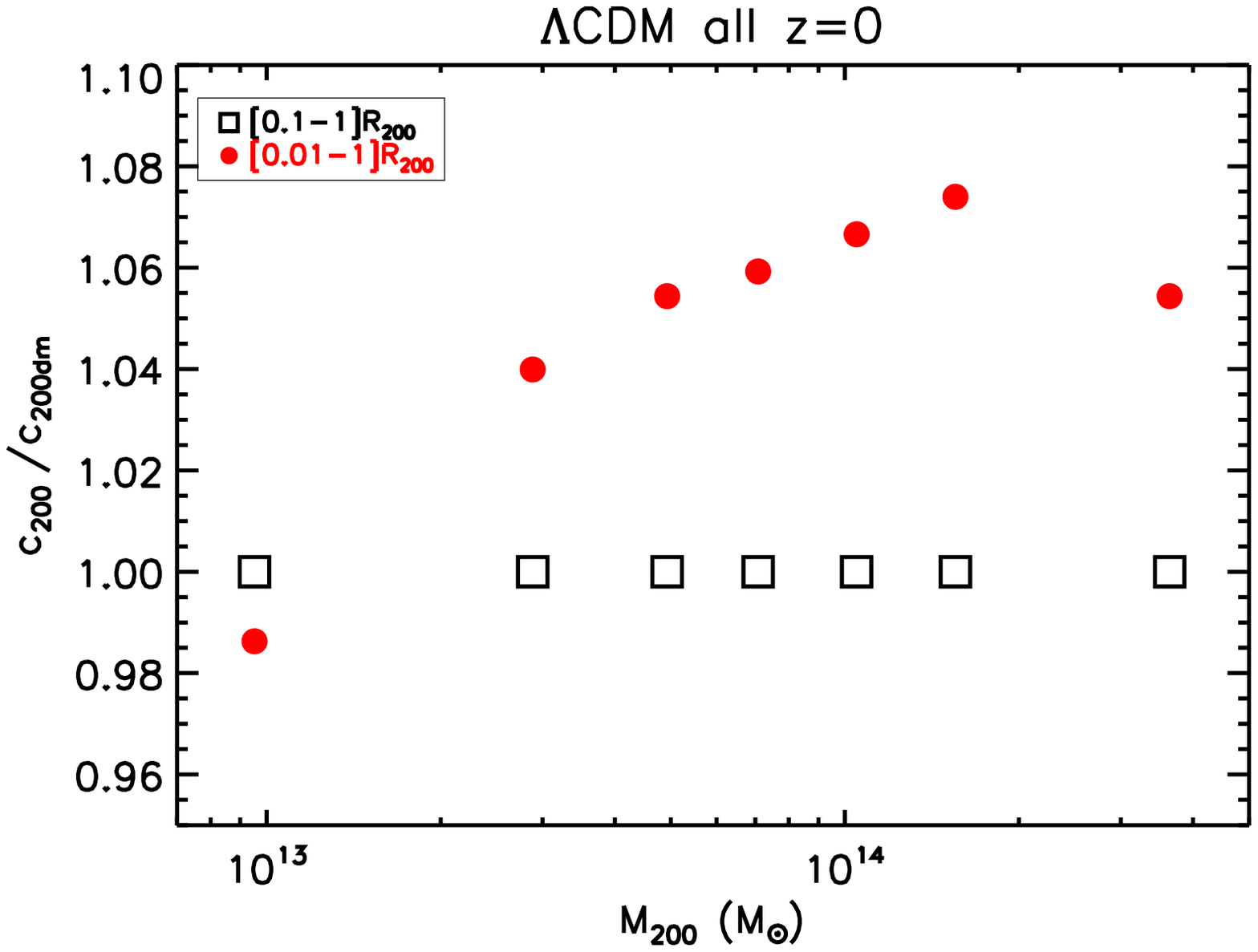,width=0.50\textwidth}
 \epsfig{figure=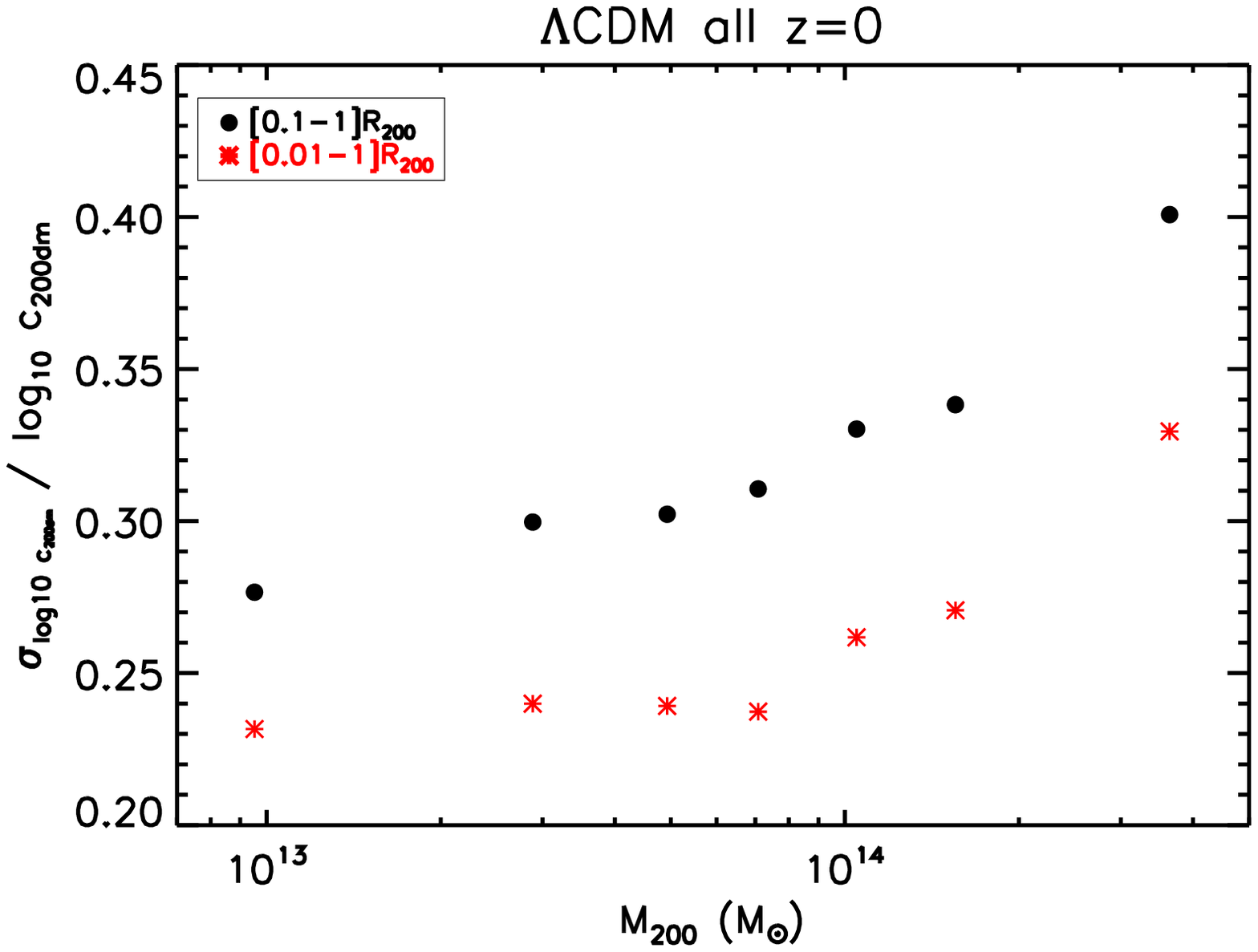,width=0.50\textwidth}
}
\caption{(Left panel) Ratio between $c_{200dm}$ evaluated from equation (\ref{c_fit}) by fitting equation (\ref{NFW_c-M}) in the range [$0.1-1$]$R_{200}$ (black squares) and $c_{200dm,fit}$ evaluated by fitting in the range [$0.01-1$]$R_{200}$ (red points) and $c_{200dm}$ evaluated by fitting in the range [$0.1-1$]$R_{200}$, for the complete sample of the $\Lambda$CDM model at $z=0$. (Right panel) Relative error in ${\rm{log_{10}}}  c_{200dm}$ for the [$0.1-1$]$R_{200}$ fit (black points) and the [$0.01-1$]$R_{200}$ fit (red stars) for the complete sample of the $\Lambda$CDM model at $z=0$.}
\label{fitting_radius}
\end{figure*}

\noindent As a check, for the complete sample of the $\Lambda$CDM model at $z=0$ we also evaluate $c_{200dm}$ by fitting the NFW profile equation (\ref{NFW_c-M}) in the range [$0.01-1$]$R_{200}$. For low mass objects in our sample, the value of $0.01R_{200}$ is close to the limit of the force resolution of the simulation, which is $7.5 \ {\rm{kpc}} \ h^{-1}$ at $z=0$. In Fig \ref{fitting_radius} we show the results for the values of $c_{200dm}$ and the relative error in ${\rm{log_{10}}}  c_{200dm}$ in each mass bin. We see that, by fitting in the range [$0.01-1$]$R_{200}$, we obtain concentrations up to $10\%$ higher than by fitting in the [$0.1-1$]$R_{200}$ range. Moreover, by fitting including the inner regions in the fit, the intrinsic scatter in concentration is lower by about $5\%$. These trends are almost independent of mass. Since we want to compare the dark matter only runs with the hydrodynamical ones, and in the hydrodynamical runs we do not completely resolve the baryonic physics on very small scales, we will take a conservative approach and fit in the range [$0.1-1$]$R_{200}$. Still, it is important to know what happens if we consider also the inner regions, in particular if we want to compare our results with the ones in literature.

With the mean and rms deviation of ${\rm{log_{10}}} c_{200dm}$ in each bin at hand, we fit, for the complete, relaxed and super-relaxed samples, the binned $c-M$ relation using

\begin{equation}
{\rm{log_{10}}}  c_{200}={\rm{log_{10}}}  A + B \ {\rm{log_{10}}}  \left( \frac{M_{200}}{10^{14} \ {\rm{M_{\odot}}}} \right) \ ,
\label{c-M}
\end{equation}

\noindent where ${\rm{log_{10}}} c_{200}$ and $M_{200}$ are the mean values in each bin. For the error on the mean of ${\rm{log_{10}}} c_{200dm}$ in each bin, $\sigma_{\bar{c}}$, we use the rms deviation of ${\rm{log_{10}}} c_{200dm}$ divided by the square root of the number of objects in the bin. For each fit we also define

\begin{equation}
\chi^2 = \sum_{i=1}^{N_{bins}} \left( \frac{{\rm{log_{10}}}  c_{200_{i}} - {\rm{log_{10}}}  c_{200_{fit}}}{\sigma_{\bar{c}_{i}}} \right)^2 \ 
\end{equation}

\noindent and evaluate the reduced chi-squared $\tilde{\chi}^2$, {\it{i.e.}} ${\chi}^2$ divided by the number of degrees of freedom. We list the best fit values $A$ and $B$ for each sample, along with the corresponding standard errors and $\tilde{\chi}^2$, in Table \ref{tab_super-relaxed_parameters_fit_comparison_dm}. We see that, compared to the complete sample, the normalization $A$ increases by about $15\%$ for the relaxed sample and by about $25\%$ for the super-relaxed sample, while the slope $B$ does not change significantly, even if excluding unrelaxed objects results in a shallower slope. The values of $\tilde{\chi}^2$ indicate that equation (\ref{c-M}) is a good parametrization of the $c-M$ relation in logarithmic scale.

\begin{table}
\caption{Best-fit parameters, standard errors and reduced chi-squared $\tilde{\chi}^2$ of the $c-M$ relation equation (\ref{c-M}) for dark matter only density profile fit in the region [$0.1-1$]$R_{200}$ for the complete, relaxed and super-relaxed samples of the $\Lambda$CDM model at $z=0$.}
\begin{tabular}{|cc|cc|cc|c|}
\hline Model & $\sigma_8$ & $A$ & $\sigma_{A}$ & $B$ & $\sigma_{B}$ & $\tilde{\chi}^2$ \\
\hline $\Lambda$CDM & $0.776$ & \multicolumn{5}{|c|}{dm} \\
\hline
all & & $3.59$ & $0.05$ & $-0.099$ & $0.011$ & $0.48$ \\
relaxed & & $4.09$ & $0.05$ & $-0.092$ & $0.011$ & $0.66$ \\
super-relaxed & & $4.52$ & $0.06$ & $-0.091$ & $0.013$ & $0.76$ \\
\hline
\end{tabular}
\label{tab_super-relaxed_parameters_fit_comparison_dm}
\end{table}

\noindent In order to understand the impact of low-mass object on the $c-M$ relation, we check how the best-fit values of the $c-M$ relation change if we do not include the less massive objects. We report the results we obtain by considering only objects with $M_{200m} > 10^{14} \ {\rm{M_{\odot}}} \ h^{-1}$ in Table \ref{tab_super-relaxed_parameters_fit_comparison_dm_mass_cut}. For all three samples, we find a flatter relation than when including also low-mass objects, with larger errors on the slope. The normalizations are lower of few percentage points, while the relative errors are a factor of two higher compared to the case where low-mass objects are also considered. Moreover, we find that in this case the slope is very sensible to the dynamical state of the objects included in the sample. Thus we can conclude that the inclusion of low-mass objects is necessary to find a significant correlation between the concentration and the mass of the halos. We do not quote the reduced chi-squared in this case because just 3 mass bins are considered with 2 parameters to be fitted.

\begin{table}
\caption{Best-fit parameters and standard errors of the $c-M$ relation equation (\ref{c-M}) for dark matter only density profile fit in the region [$0.1-1$]$R_{200}$, considering only objects with $M_{200m} > 10^{14} \ {\rm{M_{\odot}}} \ h^{-1}$, for the complete, relaxed and super-relaxed samples of the $\Lambda$CDM model at $z=0$.}
\begin{tabular}{|cc|cc|cc|}
\hline Model & $\sigma_8$ & $A$ & $\sigma_{A}$ & $B$ & $\sigma_{B}$ \\
\hline $\Lambda$CDM & $0.776$ & \multicolumn{4}{|c|}{dm} \\
\hline
all & & $3.55$ & $0.09$ & $-0.087$ & $0.038$ \\
relaxed & & $3.99$ & $0.10$ & $-0.055$ & $0.042$ \\
super-relaxed & & $4.35$ & $0.13$ & $-0.010$ & $0.049$ \\
\hline
\end{tabular}
\label{tab_super-relaxed_parameters_fit_comparison_dm_mass_cut}
\end{table}

For the dark matter only profiles of the concordance $\Lambda$CDM model we also perform a logarithmic fit of equation (\ref{NFW_c-M}) without using Poissonian errors, as usually found in the literature. In this case we evaluate $c_{200}$ both from equation (\ref{c_fit}) and by directly defining $c_{200} \equiv R_{200}/r_{s}$ (using $R_{200}$ from the true mass profile), and indicate the two values as $c_{200dm,fit}$ and $c_{200dm,rec}$, respectively.

\noindent Moreover, in order to check the robustness of our fit, we perform a logarithmic fit, using Poissonian errors, of the dark matter profile times $r^2$, and a logarithmic fit, using Poissonian errors, of the dark matter profile times $r^3$.

\noindent Finally, we perform a logarithmic one-parameter fit, using Poissonian errors, of equation (\ref{NFW_c-M}) re-expressed as

\begin{eqnarray}
&& \frac{\rho_{dm}(r)}{\rho_c} = \label{NFW_one} \\
&& =\frac{200}{3}\frac{c_{200}^3}{\left[\ln (1+ c_{200}) - \frac{c_{200}}{1+c_{200}}\right]}^{   } \frac{1}{\left[ \frac{c_{200}r}{R_{200}} \right] \left[1+ \left( \frac{c_{200}r}{R_{200}} \right) \right]^{2}} \ , \nonumber
\end{eqnarray}

\noindent where the only free parameter is $c_{200}$, since $R_{200}$ is directly taken from the true mass profile. 

\noindent Only for comparison, we also evaluate the halos concentration following \cite{2011arXiv1104.5130P}, {\it{i.e.}} by solving

\begin{equation}
\frac{V_{max}}{V_{200}} = \left(  \frac{0.216 \ c}{f(c)} \right)^{1/2} \ ,
\label{vmax}
\end{equation}

\noindent where

\begin{equation}
V_{max}= {\rm{max}} \left[ \frac{GM(<r)}{r} \right]^{1/2} \ ,
\end{equation}

\begin{equation}
V_{200}=\left( \frac{GM_{200}}{R_{200}} \right)^{1/2} \ ,
\end{equation}

\noindent and, for a NFW profile, $f(c)$ is given by 

\begin{equation}
f(c) = \ln (1+c) -\frac{c}{1+c} \ .
\end{equation}

\begin{figure}
\hbox{
 \epsfig{figure=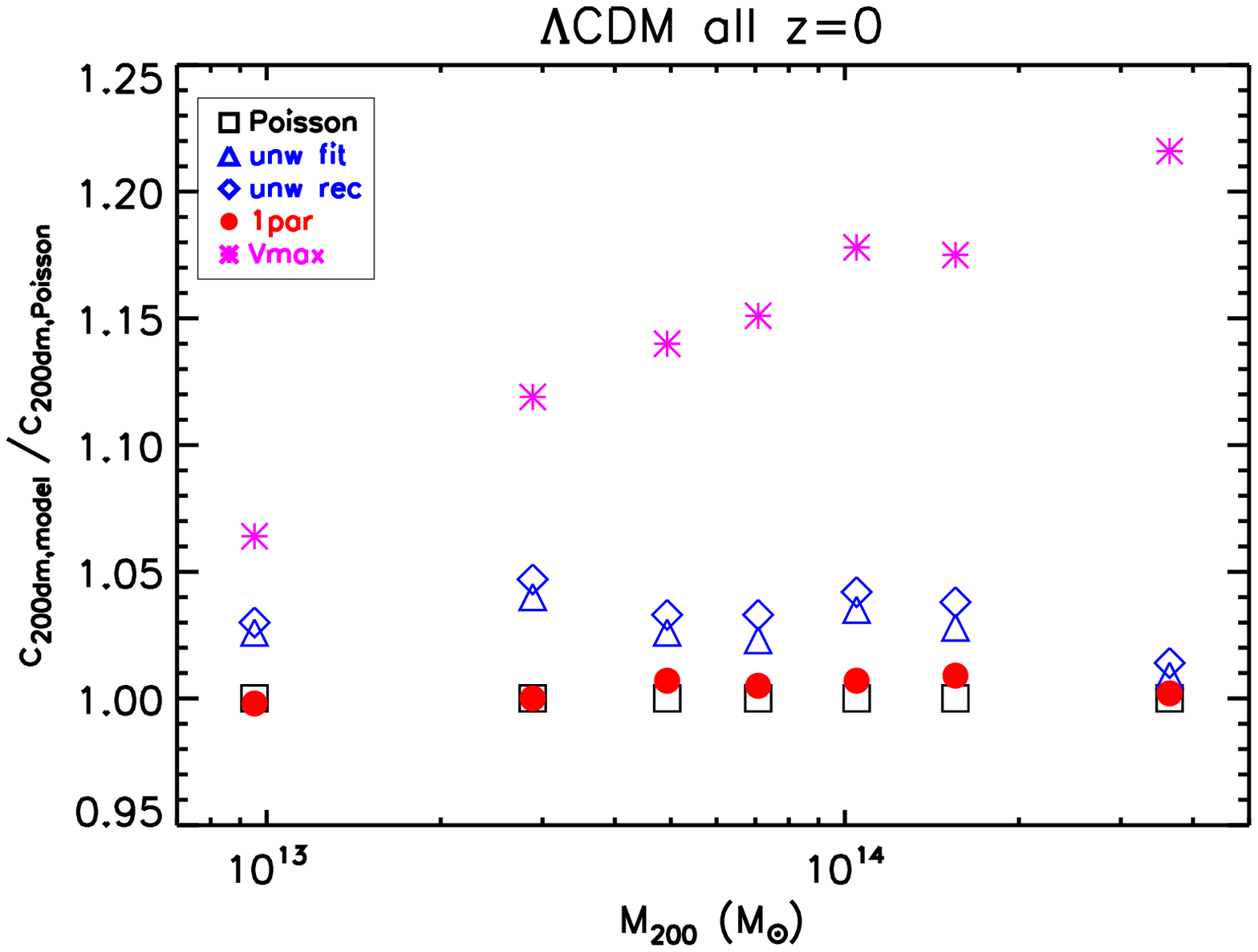,width=0.50\textwidth}
}
\hbox{
 \epsfig{figure=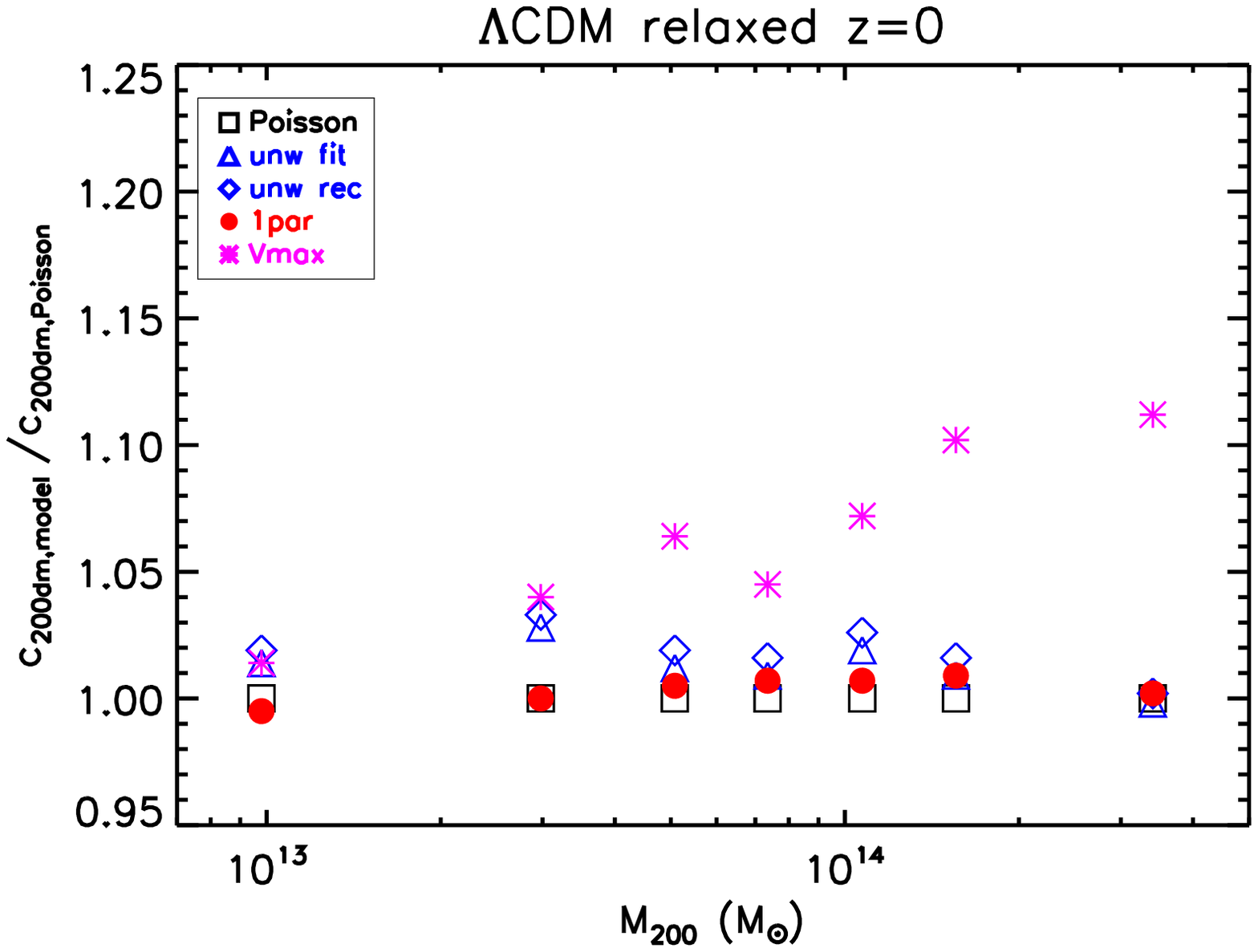,width=0.50\textwidth}
}
\hbox{
 \epsfig{figure=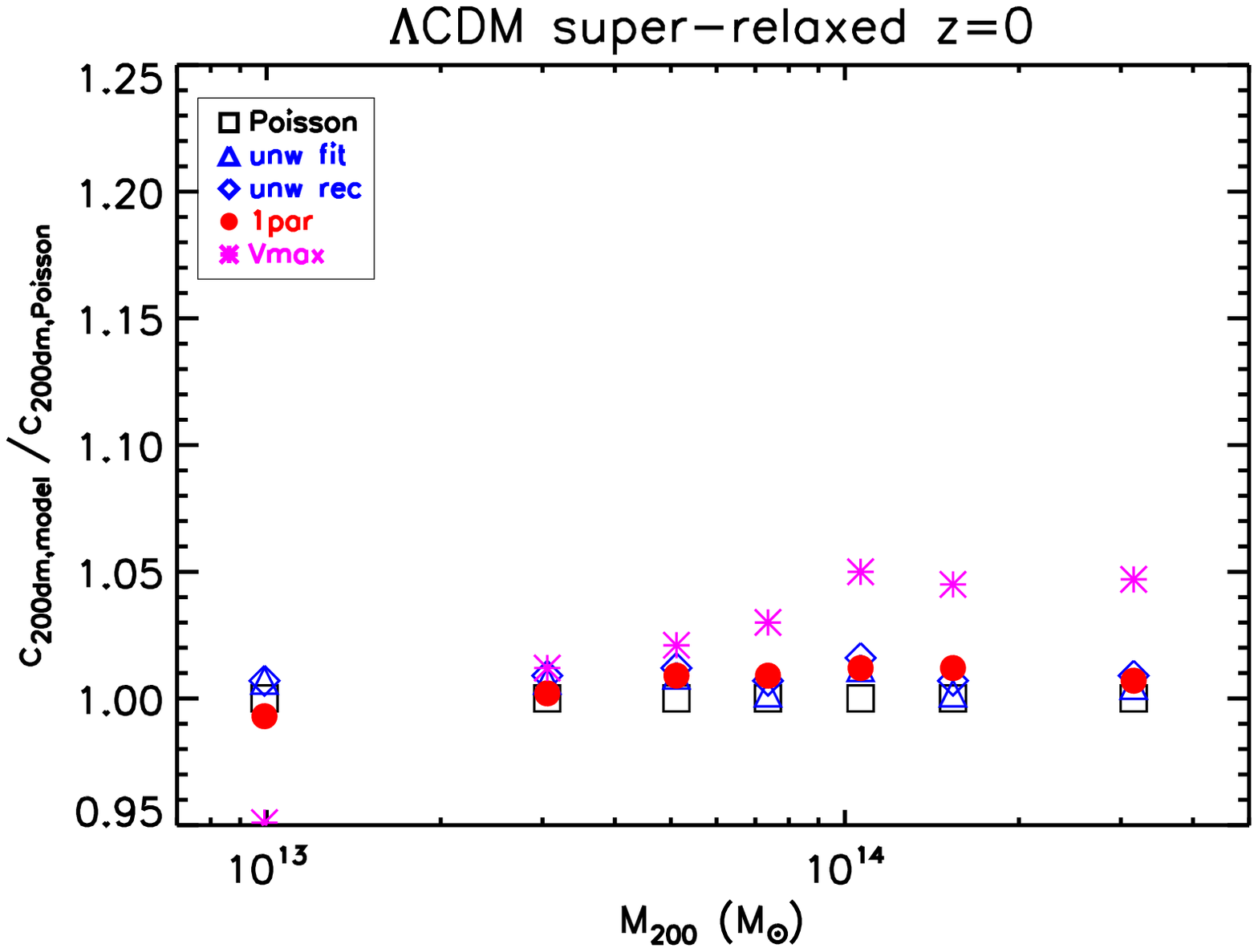,width=0.50\textwidth}
}

\caption{(Top panel) Ratio between $c_{200dm}$ (black squares), $c_{200dm,fit}$ (blue triangles), $c_{200dm,rec}$ (blue diamonds), $c_{200dm}$ found by fitting equation (\ref{NFW_one}) (red points), and $c$ recovered from equation (\ref{vmax}) (pink stars) and $c_{200dm}$ evaluated from equation (\ref{c_fit}) by fitting equation (\ref{NFW_c-M}) using Poissonian errors, for the complete sample of the $\Lambda$CDM model at $z=0$. (Middle panel) The same as in the top panel, but for the relaxed sample. (Bottom panel) The same as in the top panel, but for the super-relaxed sample.}
\label{fit_comparison}
\end{figure}

\noindent We find that the best values [$\delta$, $r_{s}$], and therefore $c_{200}$ from equation (\ref{c_fit}), found by fitting the dark matter profile times $r^2$ and the dark matter profile times $r^3$ are exactly the ones found from equation (\ref{NFW_c-M}) with Poissonian errors, while the ones from equation (\ref{NFW_c-M}) without Poissonian errors, from equation (\ref{NFW_one}), and from equation (\ref{vmax}) are somehow different. We show these differences for the complete, relaxed, and super-relaxed samples in Fig. \ref{fit_comparison}. For the complete sample, we notice from Fig. \ref{fit_comparison} that the unweighted fit tends to give values of $c_{200dm}$ slightly higher compared to the Poisson weighted fits, with $c_{200dm,rec}$ always higher than $c_{200dm,fit}$. The values of $c_{200dm}$ found from the one-parameter fit of equation (\ref{NFW_one}) are consistent with the ones found from equation (\ref{NFW_c-M}) with Poissonian errors. All these models are compatible within few percentage points, while the method using $V_{max}$ discussed in \cite{2011arXiv1104.5130P} gives systematically higher values, with a positive trend with mass. In the most massive bin, the difference between the concentrations is more than $20\%$. For the relaxed sample, the situation is similar, but the differences among the different fits are somewhat smaller. Even for the method based on equation (\ref{vmax}), which is the one that gives very diverse results, the difference is at most around $10\%$. For the super-relaxed sample, the differences among the various fits almost disappear, with the exception of the method using $V_{max}$, which however now shows differences below $5\%$ and, in the lowest mass bin, recovers values of the concentration even lower than the other methods. The method based on $V_{max}$ recovers the same values of the concentration of the usual NFW fit for super-relaxed objects, as expected because it is based on the NFW profile. But for less relaxed objects it tends to maximize the value of $V_{max}/V_{200}$ by taking into account features due to accretion/merging, thus artificially enhancing the concentration of the halos.

In Table \ref{tab_parameters_fit_comparison_dm} we show the best-fit parameters, the standard errors and the reduced chi-squared of the $c-M$ relation for the different fits discussed above. As expected, for all the three samples, the differences in the normalization and slope are limited to a few precent, with the exception of the method discussed in \cite{2011arXiv1104.5130P}. For this method we find a flatter relation, due to the fact that it predicts higher concentrations for high-mass objects compared to the usual fit of the NFW profile equation (\ref{NFW_c-M}). Moreover, also the normalization is higher (more than $15\%$ for the complete sample, more than $5\%$ for the relaxed sample, and less than $5\%$ for the super-relaxed one), given that the concentration found from equation (\ref{vmax}) is in general higher than the one found in the usual way. As already shown, the effect is stronger for the complete sample than for the relaxed and super-relaxed ones. The values of the reduced chi-squared indicate that equation (\ref{c-M}) is a good parametrization of the $c-M$ relation also for these different definitions of the concentration.

\begin{table} \small
\caption{Best-fit parameters, standard errors and reduced chi-squared $\tilde{\chi}^2$ of the $c-M$ relation equation (\ref{c-M}) for the different fits of the dark matter only density profile for the complete, relaxed and super-relaxed samples of the $\Lambda$CDM model at $z=0$.}
\begin{tabular}{|cc|cc|cc|c|}
\hline Model & $\sigma_8$ & $A$ & $\sigma_{A}$ & $B$ & $\sigma_{B}$ & $\tilde{\chi}^2$ \\
\hline $\Lambda$CDM & $0.776$ & \multicolumn{5}{|c|}{dm (all)} \\
\hline
Poisson & & $3.59$ & $0.05$ & $-0.099$ & $0.011$ & $0.48$ \\
$c_{200dm,fit}$ & & $3.69$ & $0.04$ & $-0.102$ & $0.010$ & $0.49$ \\
$c_{200dm,rec}$ & & $3.71$ & $0.04$ & $-0.101$ & $0.010$ & $0.57$ \\
one-parameter & & $3.61$ & $0.05$ & $-0.096$ & $0.011$ & $0.50$ \\
$V_{max}$ & & $4.18$ & $0.04$ & $-0.063$ & $0.008$ & $0.67$ \\
\hline $\Lambda$CDM & $0.776$ & \multicolumn{5}{|c|}{dm (relaxed)} \\
\hline 
Poisson & & $4.09$ & $0.05$ & $-0.092$ & $0.011$ & $0.66$ \\
$c_{200dm,fit}$ & & $4.14$ & $0.05$ & $-0.096$ & $0.010$ & $0.53$ \\
$c_{200dm,rec}$ & & $4.17$ & $0.05$ & $-0.095$ & $0.010$ & $0.55$ \\
one-parameter & & $4.12$ & $0.05$ & $-0.088$ & $0.011$ & $0.74$ \\
$V_{max}$ & & $4.40$ & $0.04$ & $-0.065$ & $0.008$ & $0.90$ \\
\hline $\Lambda$CDM & $0.776$ & \multicolumn{5}{|c|}{dm (super-relaxed)} \\
\hline 
Poisson & & $4.52$ & $0.06$ & $-0.091$ & $0.013$ & $0.76$ \\
$c_{200dm,fit}$ & & $4.55$ & $0.06$ & $-0.092$ & $0.013$ & $0.67$ \\
$c_{200dm,rec}$ & & $4.56$ & $0.06$ & $-0.091$ & $0.013$ & $0.66$ \\
one-parameter & & $4.56$ & $0.06$ & $-0.086$ & $0.012$ & $0.63$ \\
$V_{max}$ & & $4.69$ & $0.06$ & $-0.060$ & $0.010$ & $0.45$ \\
\hline
\end{tabular}
\label{tab_parameters_fit_comparison_dm}
\end{table}

\noindent Before moving to the hydrodynamical simulation, we sum up here our findings for the dark matter only run of the reference $\Lambda$CDM model. The intrinsic dispersion in the logarithmic values of $c_{200dm}$ for objects of similar mass is reduced by a factor of two if we limit our analysis to objects that are both relaxed and with a clear NFW-like shape of the dark matter profile. Still, at best, the intrinsic scatter is of the order of $15\%$. The more relaxed the objects in the sample, the higher the normalization $A$ of the $c-M$ relation, while the slope $B$ is almost independent of the dynamical state of the halos. This is no longer true if we focus our analysis to objects with $M_{200m} > 10^{14} \ {\rm{M_{\odot}}} \ h^{-1}$, for which the slope is shallower for the relaxed sample and almost flat for the super-relaxed one.
If we limit ourselves in fitting the dark matter profile with a NFW profile, we almost recover the same values of $c_{200dm}$, $A$, and $B$, independently of the way we treat the errors on the fit and on the number of free parameters that we fit. Things do change if we use the method discussed in \cite{2011arXiv1104.5130P}, which systematically overestimates the concentration compared to the others, in particular for high-mass objects. This results in both higher values of $A$ and $B$ for this method.
With this in mind, in the following of this paper, in particular when we study the effect of dark energy on the $c-M$ relation, we will always consider $c_{200dm}$ recovered from equation (\ref{c_fit}) from the fit of equation (\ref{NFW_c-M}) with Poissonian errors, and distinguish only between the complete and the relaxed sample, whose definition is independent of the way we find the concentration.

\subsection{Hydrodynamical runs}

\begin{figure}
\hbox{
 \epsfig{figure=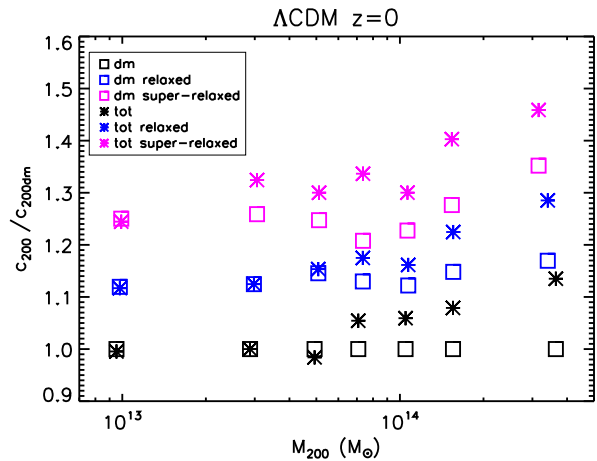,width=0.50\textwidth}
}
\caption{Ratio between $c_{200dm}$ (squares) and $c_{200tot}$ (stars) for the complete (black), relaxed (blue) and super-relaxed (pink) samples and $c_{200dm}$ for the complete sample of the $\Lambda$CDM model at $z=0$.}
\label{dm_vs_tot_comparison}
\end{figure}

\begin{figure}
\hbox{
 \epsfig{figure=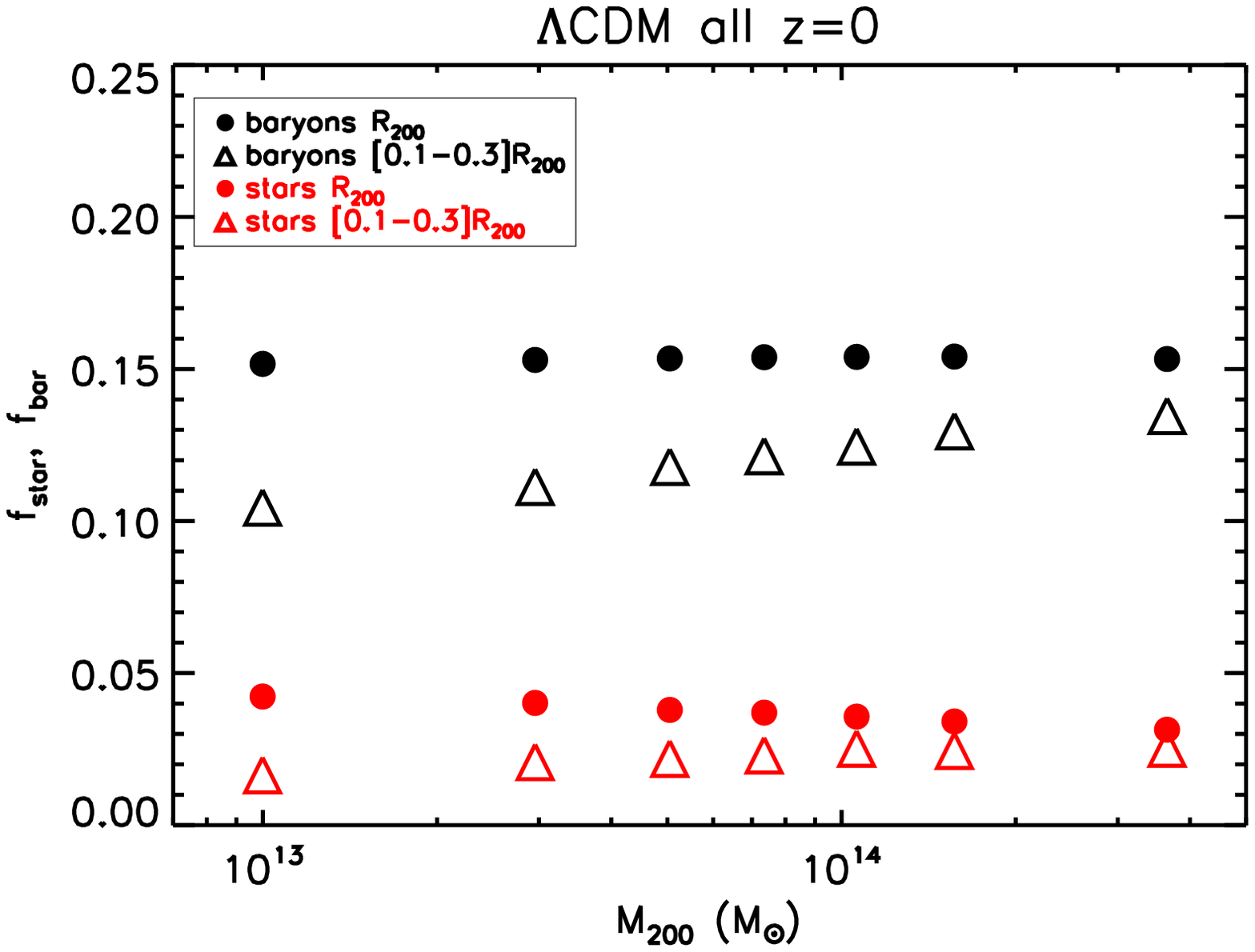,width=0.50\textwidth}
}
\hbox{
 \epsfig{figure=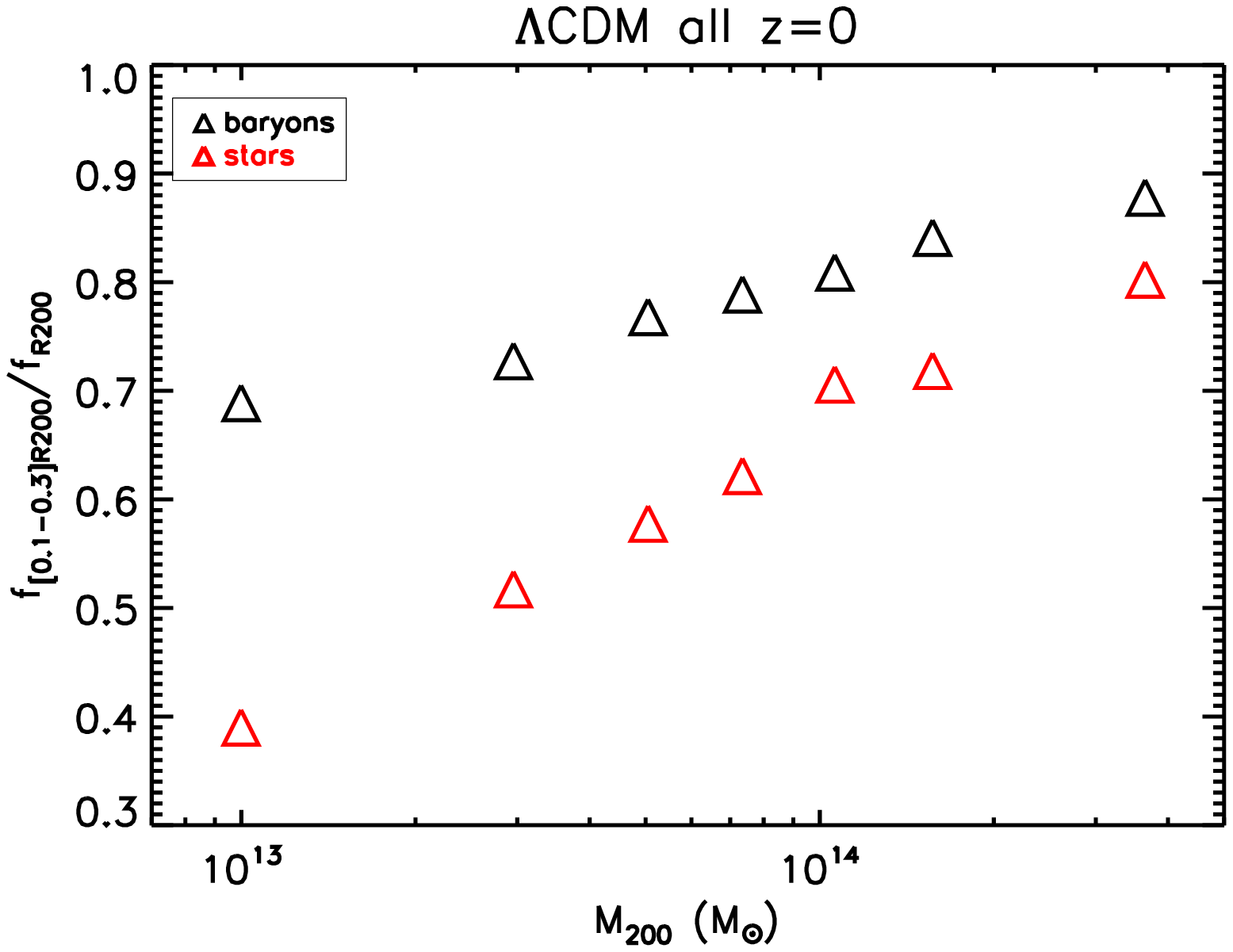,width=0.50\textwidth}
}
\caption{(Top panel) Star fraction $f_{star}$ (red) and total baryon fraction $f_{bar}$ (black) at $R_{200}$ (points) and in the $[0.1-0.3]R_{200}$ range (triangles) for the complete sample of the $\Lambda$CDM model at $z=0$. (Bottom panel) Ratio between $f_{star}$ in the range $[0.1-0.3]R_{200}$ and $f_{star}$ at $R_{200}$ (red triangles) and between $f_{bar}$ in the range $[0.1-0.3]R_{200}$ and $f_{bar}$ at $R_{200}$ (black triangles) for the complete sample of the $\Lambda$CDM model at $z=0$.}
\label{baryon_fraction_c-M}
\end{figure}

For the hydrodynamical run, we fit the total three-dimensional density profile $\rho_{tot}=\rho_{dm}+\rho_{gas}+\rho_{star}$ in the range [$0.1-1$]$R_{200}$ with equation (\ref{NFW_c-M}), using Poissonian errors $(\ln 10 \times \sqrt{n_{i}})^{-1}$ for each component (where $n_{i}$ is the number of particles of the $i$-th species) and summing them in quadrature. We indicate the total matter concentration found from equation (\ref{c_fit}) with $c_{200tot}$. We show the differences between $c_{200dm}$ and $c_{200tot}$ for the complete, relaxed and super-relaxed samples in Fig. \ref{dm_vs_tot_comparison}. We clearly see that, in all three samples, starting from objects with $M_{200m} > 10^{14} \ {\rm{M_{\odot}}} \ h^{-1}$ the concentration in the hydrodynamical run is higher than in the dark matter only case, and the effect becomes more relevant at higher masses. This effect is less pronounced in objects with $M_{200m} < 10^{14} \ {\rm{M_{\odot}}} \ h^{-1}$. Thus, the inclusion of baryons appears to affect more massive galaxy clusters than small groups. In order to check this fact, we evaluate the relative distribution of baryons and stars inside $R_{200}$ and in the range $[0.1-0.3]R_{200}$, {\it{i.e.}} the innermost part of the range in which we fit the NFW profile. We show the results in Fig. \ref{baryon_fraction_c-M}. We see that, while the total baryon fraction at $R_{200}$ is almost constant with mass, if we limit to the range $[0.1-0.3]R_{200}$ the baryon contribution to the total mass becomes more important for more massive objects. The same happens for the stars, which are known to concentrate in the internal regions of halos. Thus, in this simulation, the relative contribution of baryons and stars in the inner regions is more relevant in massive galaxy clusters than in small groups, and this fact leads to an increase of the concentration in massive objects when including baryonic physics in the simulations.

\noindent Then, we fit the $c-M$ relation equation (\ref{c-M}) as in the dark matter only case. We list the best-fit parameters, standard errors and reduced chi-squared of the $c-M$ relation for the complete, relaxed, and super-relaxed samples in Table \ref{tab_parameters_comparison_tot}. For all samples, the normalization $A$ is from $5\%$ to $10\%$ higher compared to the dark matter only case, while the slope is shallower by about $30\%$. This is expected because, as we have already seen, in the hydrodynamical runs high-mass objects are more concentrated than objects of the same mass in the dark matter only simulations. For the hydrodynamical runs the values of the reduced chi-squared are quite high. This fact seems to indicate that baryons introduce some dependence and equation (\ref{c-M}), even though it remains a good parametrization, is no more able to completely characterize.

\begin{table}
\caption{Best-fit parameters, standard errors and reduced chi-squared $\tilde{\chi}^2$ of the $c-M$ relation equation (\ref{c-M}) for dark matter only and total density profile fit in the region [$0.1-1$]$R_{200}$ for the complete, relaxed and super-relaxed samples of the $\Lambda$CDM model at $z=0$.}
\begin{tabular}{|cc|cc|cc|c|}
\hline Model & $\sigma_8$ & $A$ & $\sigma_{A}$ & $B$ & $\sigma_{B}$ & $\tilde{\chi}^2$ \\
\hline $\Lambda$CDM & $0.776$ & \multicolumn{5}{|c|}{all} \\
\hline
dm & & $3.59$ & $0.05$ & $-0.099$ & $0.011$ & $0.48$ \\
total & & $3.81$ & $0.05$ & $-0.061$ & $0.011$ & $1.69$ \\
\hline $\Lambda$CDM & $0.776$ & \multicolumn{5}{|c|}{relaxed} \\
\hline 
dm & & $4.09$ & $0.05$ & $-0.092$ & $0.011$ & $0.66$ \\
total & & $4.29$ & $0.05$ & $-0.064$ & $0.011$ & $1.54$ \\
\hline $\Lambda$CDM & $0.776$ & \multicolumn{5}{|c|}{super-relaxed} \\
\hline 
dm & & $4.52$ & $0.06$ & $-0.091$ & $0.013$ & $0.76$ \\
total & & $4.89$ & $0.06$ & $-0.062$ & $0.011$ & $1.34$ \\
\hline
\end{tabular}
\label{tab_parameters_comparison_tot}
\end{table}

\subsection{The dependence upon redshift} 

\begin{figure}
\hbox{
 \epsfig{figure=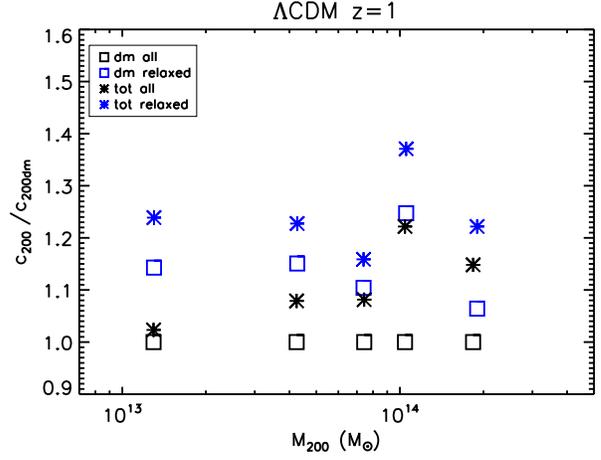,width=0.50\textwidth}
}
\caption{Ratio between $c_{200dm}$ (squares) and $c_{200tot}$ (stars) for the complete (black) and relaxed (blue) samples and $c_{200dm}$ for the complete sample of the $\Lambda$CDM model at $z=1$.}
\label{dm_vs_tot_comparison_z1}
\end{figure}

Finally, in order to study the evolution with redshift of the $c-M$ relation, which is of fundamental importance if we want to distinguish among different cosmological models, we also consider objects at $z=0.5$ and $z=1$. We show the differences between $c_{200dm}$ and $c_{200tot}$ for the complete and relaxed sample of the $\Lambda$CDM model at $z=1$ in Fig. \ref{dm_vs_tot_comparison_z1}. Even if the trend with mass is less clear than at $z=0$, still we can see that, already at $z=1$, $c_{200tot}$ is greater than $c_{200dm}$ in both samples.

\noindent When we consider also objects at $z=0.5$ and $z=1$, we fit a generalized form of the $c-M$ relation equation (\ref{c-M}) with an explicit redshift dependence, namely
 
\begin{eqnarray}
\label{c-M_z}
&& {\rm{log_{10}}}  c_{200}= \\
&& {\rm{log_{10}}}  A + B \ {\rm{log_{10}}}  \left( \frac{M_{200}}{10^{14} \ {\rm{M_{\odot}}}} \right) + C \ {\rm{log_{10}}}  (1+z) \ . \nonumber
\end{eqnarray}

\noindent We can perform this fit in two ways. Either we keep the best-fit values $A_{0}$ and $B_{0}$ found at $z=0$ fixed and fit only $C$ in equation (\ref{c-M_z}), or we perform a three-parameter fit by keeping $A$, $B$ and $C$ free. We report the results of the different fits both for the complete and relaxed samples of the dark matter only and hydrodynamical simulations in Table \ref{tab_parameters_c-M-z}. We note that, in all cases, the redshift dependence is negative, meaning that objects of a given mass have lower concentration at higher redshift. This is expected, because $c \sim \rho_{c}^{-1/3}$ and the critical density drops with decreasing redshift. If we leave the normalization and the slope free, we see that $A$ changes at most of few percentage points, while $B$ can vary significantly. The redshift dependence seems to be insensitive to both the way in which the fit is performed and the dynamical state of the halos, while it is different for dark matter only and total concentration, being steeper in the former case and shallower in the latter. In general, the reduced chi-squared of the fit of equation (\ref{c-M_z}) is rather high, in particular for the hydrodynamical runs. This can be an indication that the redshift dependence we are considering is somehow too simple to fully reproduce the redshift evolution of the $c-M$ relation, and that the presence of baryons makes this evolution more complex. In the following sections, we will constrain the normalization and slope at $z=0$, and then study the redshift evolution keeping $A$ and $B$ fixed.

\begin{table*}
\caption{Best-fit parameters, standard errors and reduced chi-squared $\tilde{\chi}^2$ of the $c-M$ relation equation (\ref{c-M_z}) for dark matter only and total density profile fit in the region [$0.1-1$]$R_{200}$ for the complete and relaxed samples of the $\Lambda$CDM model at $z=0$, $z=0.5$ and $z=1$ both fixing $A$ and $B$ at the best-fit values at $z=0$ and keeping all the parameters free.}
\begin{tabular}{|cc|cc|cc|cc|c|}
\hline Model & $\sigma_8$ & $A$ & $\sigma_{A}$ & $B$ & $\sigma_{B}$ & $C$ & $\sigma_{C}$ & $\tilde{\chi}^2$ \\
\hline $\Lambda$CDM & $0.776$ & \multicolumn{7}{|c|}{dm (all)} \\
\hline
fixed & & $3.59$ & --- & $-0.099$ & --- & $-0.33$ & $0.02$ & $1.68$ \\
free & & $3.63$ & $0.04$ & $-0.077$ & $0.008$ & $-0.32$ & $0.03$ & $1.41$ \\
\hline $\Lambda$CDM & $0.776$ & \multicolumn{7}{|c|}{dm (relaxed)} \\
\hline
fixed & & $4.09$ & --- & $-0.092$ & --- & $-0.31$ & $0.02$ & $1.22$ \\
free & & $4.13$ & $0.05$ & $-0.080$ & $0.008$ & $-0.31$ & $0.03$ & $1.24$ \\
\hline $\Lambda$CDM & $0.776$ & \multicolumn{7}{|c|}{total (all)} \\
\hline 
fixed & & $3.81$ & --- & $-0.061$ & --- & $-0.26$ & $0.02$ & $2.40$ \\
free & & $3.81$ & $0.04$ & $-0.046$ & $0.007$ & $-0.25$ & $0.03$ & $2.40$ \\
\hline $\Lambda$CDM & $0.776$ & \multicolumn{7}{|c|}{total (relaxed)} \\
\hline 
fixed & & $4.29$ & --- & $-0.064$ & --- & $-0.26$ & $0.02$ & $1.97$ \\
free & & $4.27$ & $0.05$ & $-0.071$ & $0.007$ & $-0.26$ & $0.03$ & $2.15$ \\
\hline
\end{tabular}
\label{tab_parameters_c-M-z}
\end{table*}

\section{Model-independent concentration} \label{nomodel}

\begin{figure}
\hbox{
 \epsfig{figure=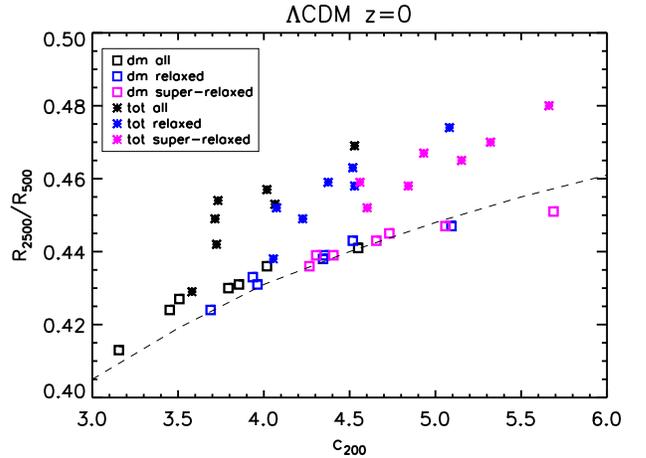,width=0.50\textwidth}
}
\caption{Comparison between $c_{200dm}$ (squares) and $c_{200tot}$ (stars) and $R_{500}/R_{2500}$ for the complete (black), relaxed (blue) and super-relaxed (pink) samples of the $\Lambda$CDM model at $z=0$. The black dashed-line represents the prediction for a NFW profile.}
\label{non_parametric}
\end{figure}

All the fits we used in the previous sections are based on some assumptions that we made on the shape of the density profile. In principle, it should be useful to have a model-independent proxy for the concentration found from equation (\ref{c_fit}). Given the usual definition of the concentration, a natural quantity could be the ratio between two physical radii, characterized by a given overdensity. As already pointed out in \cite{2010MNRAS.405.2161D}, a good choice should be $R_{2500}/R_{500}$ since these radii are commonly used in X-ray observations. For each object, we evaluate the ratio $R_{2500}/R_{500}$ taking $R_{2500}$ and $R_{500}$ directly from the true mass profile, without any assumption on the density profile. We bin the objects in mass as in the previous sections. The results for the complete, relaxed and super-relaxed samples, are shown in Fig. \ref{non_parametric}. We note is that, for a given sample, the variation of $c_{200}$ is more than $30\%$ while the variation of $R_{2500}/R_{500}$ is less than $10\%$. This means that the ratio between $R_{2500}$ and $R_{500}$ is less dependent on the mass of the halo when compared to $c_{200}$. We also notice that objects with higher concentration, {\it{i.e.}} with higher $R_{200}/r_{s}$, also have higher $R_{2500}$/$R_{500}$. This result confirms the ones in \cite{2008MNRAS.390L..64D}, where a weak positive trend of $R_{500}/R_{2500}$ with $M_{500}$ was found. Finally, there is a clear dichotomy between the dark matter only and the total profile fits. At a given concentration, when including baryons, the ratio $R_{2500}$/$R_{500}$ is more than $5\%$ higher with compared to the dark matter only case. Moreover, in the dark matter only case, the relation between $R_{2500}$/$R_{500}$ and $c_{200}$ closely follows the one predicted by a NFW profile, indicating that the halos are well described by this profile. The different behaviour of the halos extracted from the hydrodynamical run suggests that, when we consider the total (dark matter plus baryons) density profile, the NFW profile is no longer a good approximation to the real profile.

\section{Comparison with other works} \label{comparison}

In this section, we compare with previous works our results on the dark matter concentration obtained by fitting equation (\ref{NFW_c-M}) in the range [$0.1-1$]$R_{200}$. We use objects at $z=0$, $z=0.5$ and $z=1$ from the dark matter only simulations of the concordance $\Lambda$CDM model normalized with WMAP3 data. We recall that, by fitting the $c-M$ relation equation (\ref{c-M_z}), we find that more massive objects are less concentrated than less massive ones and that objects at high redshift are less concentrated than objects at $z=0$. Moreover, relaxed objects are more concentrated compared to the complete sample. All these findings qualitatively confirm what can be found in literature. For the sake of comparison with literature, some numbers are summarized in Table \ref{tab_comparison}, where we have adjusted the quoted normalization $A$ to our pivot mass $10^{14} \ \rm{M_{\odot}}$ using the corresponding quoted slope $B$. 

\begin{table*}
\caption{Best-fit parameters of the $c-M$ relation, comparison with other works. See text for details.}
\begin{tabular}{|cccc|ccc|ccc|}
\hline Reference & $h$ & $\Omega_{m}$ & $\sigma_8$ & \multicolumn{3}{|c|}{dm (all)} & \multicolumn{3}{|c|}{dm (relaxed)} \\
\hline & & & & $A$ & $B$ & $C$ & $A$ & $B$ & $C$ \\
\hline
This work (fixed $A$ and $B$) & $0.704$ & $0.268$ & $0.776$ & $3.59$ & $-0.099$ & $-0.33$ & $4.09$ & $-0.092$ & $-0.31$ \\
This work (free $A$ and $B$) & $0.704$ & $0.268$ & $0.776$ & $3.63$ & $-0.077$ & $-0.32$ & $4.13$ & $-0.080$ & $-0.31$ \\
\cite{2008MNRAS.391.1940M} & $0.71$ & $0.268$ & $0.90$ & $4.55$ & $-0.119$ & --- & $5.31$ & $-0.104$ & --- \\
 & $0.73$ & $0.238$ & $0.75$ & $3.60$ & $-0.088$ & --- & $4.13$ & $-0.083$ & --- \\
 & $0.72$ & $0.258$ & $0.796$ & $3.84$ & $-0.110$ & --- & $4.46$ & $-0.098$ & --- \\
\cite{2007MNRAS.381.1450N} & $0.73$ & $0.25$ & $0.9$ & $4.85$ & $-0.11$ & --- & $5.45$ & $-0.10$ & --- \\
\cite{2008MNRAS.390L..64D} & $0.742$ & $0.258$ & $0.796$ & $4.06$ & $-0.097$ & --- & $4.81$ & $-0.092$ & --- \\
 & $0.742$ & $0.258$ & $0.796$ & $4.23$ & $-0.084$ & $-0.47$ & $4.85$ & $-0.091$ &$-0.44$ \\
 \cite{2011arXiv1104.5130P} & $0.70$ & $0.27$ & $0.82$ & $5.31$ & $-0.074$ & --- & $5.55$ & $-0.08$ & --- \\
\hline
\end{tabular}
\label{tab_comparison}
\end{table*}

\begin{figure*}
\hbox{
 \epsfig{figure=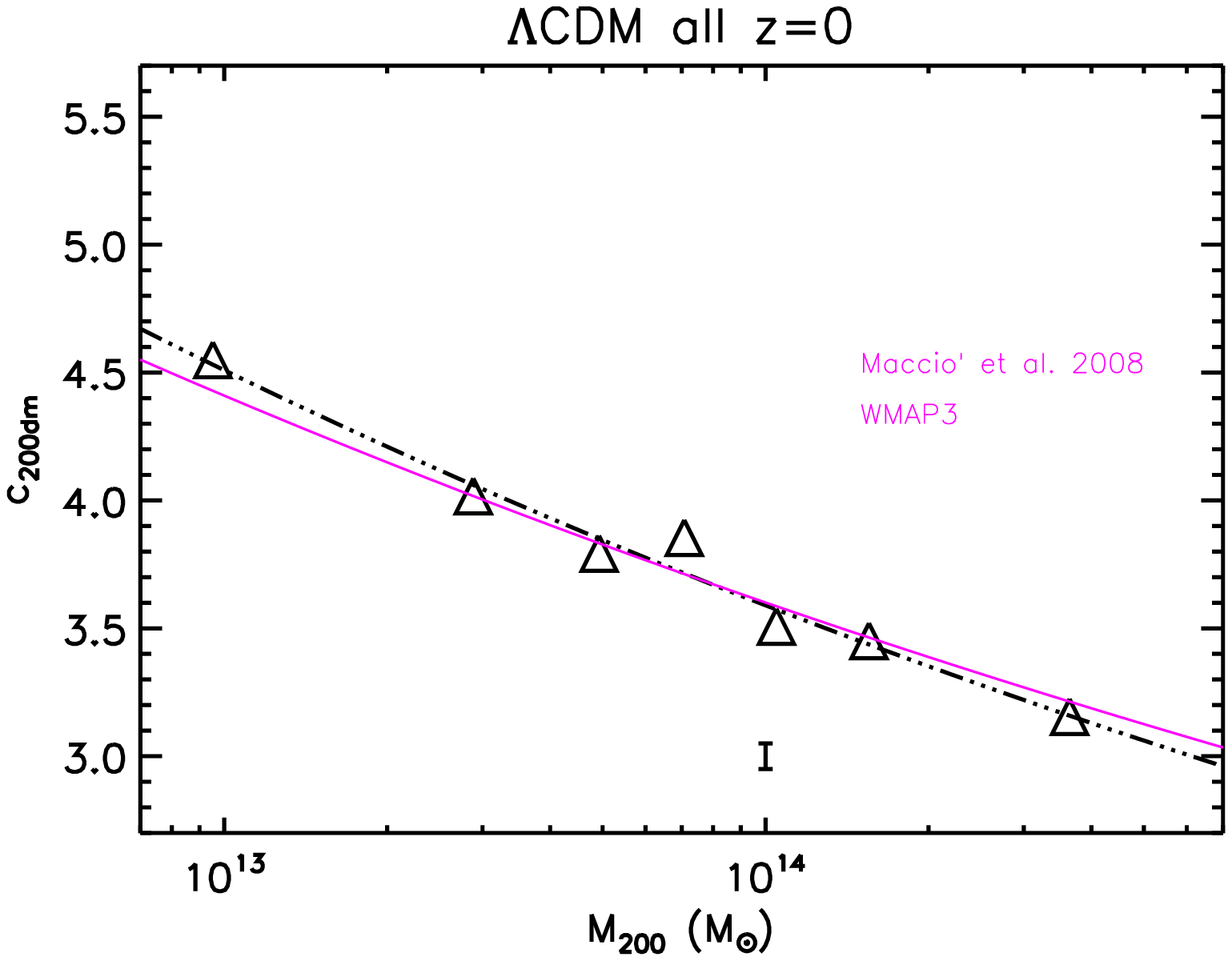,width=0.50\textwidth}
 \epsfig{figure=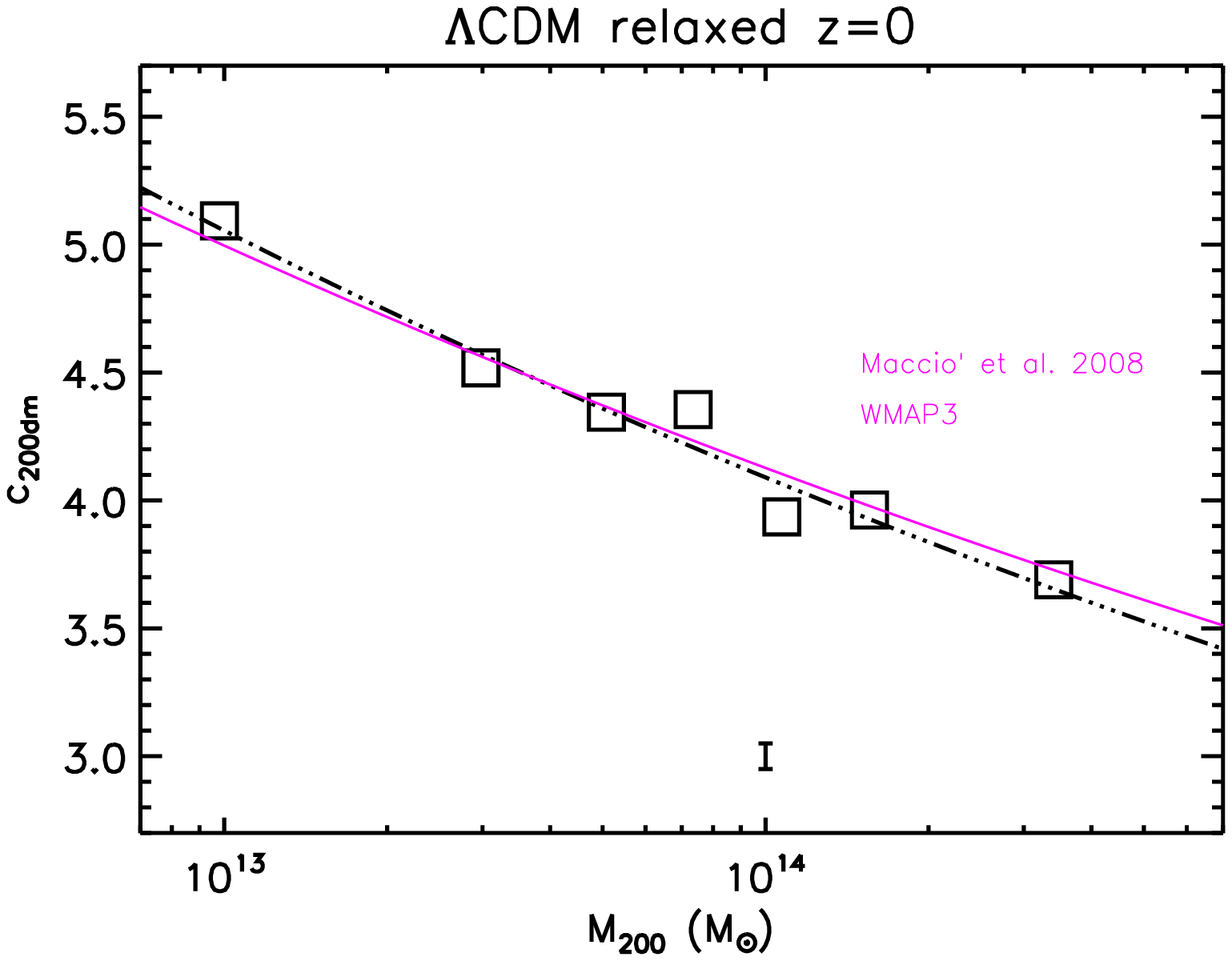,width=0.50\textwidth}
}
\caption{(Left panel) The values of $c_{200dm}$ for the complete sample of the $\Lambda$CDM cosmology at $z=0$. The black dotted-dashed line is our best-fit of $c-M$ relation equation (\ref{c-M}) and the vertical black bar is the error on the normalization as listed in Table \ref{tab_comparison}. The pink solid line is the best-fit relation of Macci\`o et al. (2008) for the WMAP3 cosmology. (Right panel) The same as in the left panel, but for the relaxed sample.}
\label{c_dm-M_comparison}
\end{figure*}

\noindent \cite{2008MNRAS.391.1940M} make a comparison of the $c-M$ relation for all and relaxed objects in different $\Lambda$CDM cosmologies, namely the ones obtained using the parameters coming from WMAP1, WMAP3 and WMAP5. They fit the dark matter profile with a NFW profile. They span the mass range $10^{10} \ {\rm{M_{\odot}}} \ h^{-1} \lesssim M \lesssim 10^{15} \ {\rm{M_{\odot}}} \ h^{-1}$ and bin the objects in mass bins of $0.4$ dex width. WMAP3 and WMAP5 data suggest a lower matter density $\Omega_{m}$ and a lower power-spectrum normalization $\sigma_{8}$ than WMAP1 ones. This means that halos of a given mass form later, and thus should be less concentrated. And indeed this is what they find, as it can be seen from Table \ref{tab_comparison}. The higher $\sigma_{8}$ and $\Omega_{m}$, the higher the normalization and the steeper the slope of the $c-M$ relation. With respect to their results for the WMAP3 cosmology, we find the same values of the normalization both for the complete and relaxed samples, with a slightly steeper slope. We show this excellent agreement in Fig. \ref{c_dm-M_comparison}.

\noindent In a more recent paper, \cite{2011MNRAS.411..584M} fit the $c-M$ relation using a formula similar to equation (\ref{c-M}), but incorporating the redshift dependence by letting $A$ and $B$ to be functions of redshift themselves. For the $c_{vir}-M_{vir}$ relation, at $z=0$ they are able to reproduce the results of \cite{2008MNRAS.391.1940M} with differences of the order of few percentage points.

\noindent \cite{2007MNRAS.381.1450N} make a comparison, using halos extracted from the Millennium Simulation (MS) \citep{2005Natur.435..629S}, between the complete sample and relaxed objects only. They fit the dark matter profile with a NFW profile in the [$0.05-1$]$R_{vir}$ range, without Poisson weighting. They span the mass range $10^{12} \ {\rm{M_{\odot}}} \ h^{-1} \lesssim M \lesssim 10^{15} \ {\rm{M_{\odot}}} \ h^{-1}$ and bin the objects in mass. They find that relaxed objects have more than $10\%$ higher normalization compared to the whole sample, while the slope is $10\%$ shallower. We find the same trend, but there is a discrepancy in the absolute numbers, both in the complete and relaxed samples. Indeed, they find normalizations that are more than $30\%$ higher than ours, which can be explained with the higher $\sigma_8$ (but also with the different radial range over which they fit, see Fig. \ref{fitting_radius} in Sect. \ref{fitting}), and a $10\%$ steeper slopes.

\noindent \cite{2008MNRAS.390L..64D} make a comparison with \cite{2007MNRAS.381.1450N} using the parameters suggested by WMAP5 (CMB only) data. They fit the dark matter profile with both a NFW and an Einasto profile in the [$0.05-1$]$R_{vir}$ range. We discuss the results for the NFW profile case. They span the mass range $10^{11} \ {\rm{M_{\odot}}} \ h^{-1} \lesssim M \lesssim 10^{15} \ {\rm{M_{\odot}}} \ h^{-1}$ and bin the objects in mass. In comparison with \cite{2007MNRAS.381.1450N}, they find lower normalizations by about $15\%$ due to a lower value of $\sigma_{8}$. They also fit the $c-M$ relation taking into account the redshift evolution from $z=0$ to $z=2$, both for the complete sample and the relaxed objects only. They find a stronger dependence on redshift than what we find, both for the whole sample and relaxed systems. A possible explanation is that we fit in the range between $z=0$ and $z=1$, while they reach $z=2$. If the redshift evolution is not constant with redshift, but it is weaker at low redshift, this could be a possible explanation for the different slope we find. \cite{2008MNRAS.390L..64D} also find that the concentration of their halos is lower than the one inferred from X-ray observations and ascribed that fact to the effect of baryon physics that was missing in their simulations. However, in \cite{2010MNRAS.405.2161D} it is shown that even including baryon physics in the simulations they cannot reproduce both observed concentrations and stellar fraction in galaxy groups and clusters. In general, at $z=0$, when including metals and AGN feedback, they find lower concentrations compared to the dark matter only case, and the effect is more relevant in low-mass objects. When including only primordial cooling, they find higher concentrations compared to the dark matter only case, qualitatively in agreement with what we find in Fig. \ref{dm_vs_tot_comparison}.

\noindent \cite{2011arXiv1104.5130P}, using the Bolshoi simulation \citep{2011ApJ...740..102K}, evaluate the concentration of the halos using equation (\ref{vmax}), binning the objects in $V_{max}$ and fitting a $c-\sigma(M,z)$ relation, where $\sigma(M,z)$ is the linear rms fluctuation of density field on the scale $M$. They find that the concentration $c(\sigma)$ has a nearly universal U-shaped profile, with some small dependence on redshift and cosmology. They also provide a fit of equation (\ref{c-M}) for all their halos at $z=0$. They find a higher normalization and a shallower slope in comparison with other works, as we do when we use equation (\ref{vmax}), instead of equation (\ref{c_fit}), to evaluate the concentration. For relaxed halos, selected by $V_{max}$, they find a $5\%$ higher normalization, as we do. Moreover, they find that the differences in concentrations for relaxed halos selected by $V_{max}$ are higher for high-mass objects than for low-mass ones.

We conclude this section noting that, when the values of the cosmological parameters are similar, our findings about the $c-M$ relation in the reference $\Lambda$CDM model are in good agreement with what already found in literature. So we can safely rely on our $\Lambda$CDM model as a reference, when comparing the impact of different dark energy models on the $c-M$ relation.

\section{Dark energy models: results on the dark matter profiles} \label{dark_matter}

From now on, we start to compare the $c-M$ relation in the $\Lambda$CDM cosmology with the ones derived for the other cosmological models under investigation. The $c-M$ relation for galaxy clusters extracted from dark matter only simulations of different dark energy models, including RP and SUGRA, has been studied in \cite{2004A&A...416..853D}. They fit a formula similar to equation (\ref{c-M_z}), but keeping $C$ fixed to $-1$. They find that, when the same $\sigma_{8}$ is used for all the models, the $c-M$ relations for dark energy cosmologies have higher normalizations compared to $\Lambda$CDM, depending on the ratio between the growth factors through

\begin{equation}
A_\mathrm{DE} \rightarrow A_\mathrm{\Lambda CDM} \frac{D_\mathrm{+,DE}(z_{coll})}{D_\mathrm{+,\Lambda CDM}(z_{coll})} \ ,
\label{c_dark_energy}
\end{equation}

\noindent where the collapse redshifts $z_{coll}$ are evaluated following the prescriptions of \cite{2001ApJ...554..114E}. When $\sigma_{8}$ values are normalized to CMB data, as we do in this work, dark energy cosmologies have lower normalizations compared to $\Lambda$CDM. We find that, in order to recover the values of the normalization they quote in this case, equation (\ref{c_dark_energy}) should be multiplied by the ratio between the values of $\sigma_{8}$, {\it{i.e.}} $\sigma_{8,\mathrm{DE}} / \sigma_\mathrm{8,\mathrm{\Lambda CDM}}$. This fact goes in the same direction as what found in \cite{2008MNRAS.391.1940M}, where models with higher $\sigma_{8}$ also have a higher normalization of the $c-M$ relation. 

We recall that to obtain the concentration we fit equation (\ref{NFW_c-M}) in the range [$0.1-1$]$R_{200}$ using Poissonian errors and adopting the best-fit parameters to obtain $c_{200}$ from equation (\ref{c_fit}). Then we bin the objects in groups of 200 halos, starting from less massive ones, and also define a relaxed sample by taking the relaxed objects inside each bin. Finally, we fit the binned $c-M$ relation with equation (\ref{c-M}). We begin the comparison in this section with the dark matter only runs at $z=0$, while in the following section we will study the hydrodynamical runs, also at higher redshifts. 

\begin{figure}
\hbox{
 \epsfig{figure=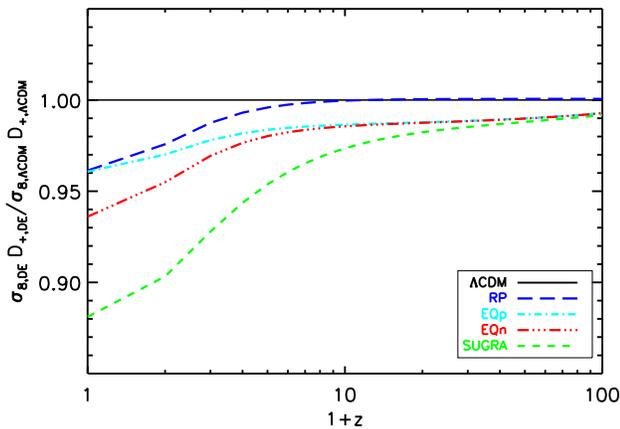,width=0.50\textwidth}
}
\caption{Ratio between the value of $\sigma_{8} D_{+}$ for the $\Lambda$CDM (black), RP (blue), EQp (cyan), EQn (red), and SUGRA (green) cosmologies and the corresponding value for $\Lambda$CDM as a function of redshift.}
\label{sigma_8_D_ratio}
\end{figure}

In Table \ref{tab_parameters_core_cut_dm} we summarize the best-fit parameters, the standard errors and the reduced chi-squared of the $c-M$ relation equation (\ref{c-M}) for the five cosmological models here considered, both for the complete and relaxed samples. For the complete sample, the differences in the normalization $A$ between $\Lambda$CDM and the other cosmological models are less than $10\%$, with EQn being the only model having a higher normalization. The slope $B$ is within $5\%$ of the $\Lambda$CDM value for all the models with the exception of EQn again, which shows a $30\%$ flatter slope. For the slope the differences among the models, excluding EQn, are smaller than the standard errors, while for the normalization these differences are significant. If we limit ourselves to the best-fit values, given that the slopes are almost identical and that all the cosmological parameters except $\sigma_{8}$ are fixed, we expect that the normalizations should follow the values of $\sigma_{8}$, {\it{i.e.}} the higher $\sigma_{8}$ the higher the normalization \citep[see][]{2008MNRAS.391.1940M}, and $D_{+}$, {\it{i.e.}} the higher $D_{+}$ at $z_{coll}$ the higher the normalization \citep[see][]{2004A&A...416..853D}. The quantity controlling the normalization is thus expected to be $\sigma_{8} D_{+}(z_{coll})$. We plot the ratio between the value of $\sigma_{8} D_{+}$ for a given dark energy model and the one for $\Lambda$CDM as a function of redshift in Fig. \ref{sigma_8_D_ratio}. Independently of the precise definition of $z_{coll}$, the cosmological model with the highest value of this quantity is $\Lambda$CDM, followed by RP, EQp, EQn, and SUGRA. We do expect the normalizations of the $c-M$ relation to follow the same order, with $\Lambda$CDM having the highest and SUGRA the lowest. Instead we see that, on the one hand, EQp which has the third highest $\sigma_{8} D_{+}$ has the lowest normalization while, on the other hand, EQn which has the second lowest $\sigma_{8} D_{+}$ has the highest normalization. The relative order of $\sigma_{8} D_{+}$ and $A$ is preserved for $\Lambda$CDM, RP and SUGRA, as in \cite{2004A&A...416..853D}. 

\noindent In a recent paper, \cite{2011arXiv1112.5479B}, using $N$-body numerical simulations of a $\Lambda$CDM cosmology, find a dependence of $c_{200}$ on both $D_{+}$ and $\nu = \delta_{c} / \sigma(M,z)$. They evaluate the overall dependence of $c_{200}$ on the linear growth factor, both for their complete and relaxed samples, to be $D_{+}^{\ 0.5}$. The dependence on the linear density contrast is also considered to be a power-law, $\nu^{a}$. They find different values of $a$ for the complete and relaxed samples, namely $a=-0.35$ for the former and $a=-0.41$ for the latter. Thus, the higher the value of $D_{+}$ the higher the value of the concentration, and the lower the value of $\delta_{c}$ the higher the value of the concentration. Our findings on the EQp and EQn models do not fit in this general picture because both $D_{+}$ and $\delta_{c}$ are lower in these models than in $\Lambda$CDM, and the differences in $\delta_{c}$ are less than $1\%$.

\noindent We hint that the behaviour of EQp and EQn is linked to the redshift evolution of the effective gravitational interaction $\tilde{G}$, as pointed out in Sect. \ref{models}. In fact, in contrast with $\Lambda$CDM, RP and SUGRA, in EQ models the gravitational constant $G$ is substituted by $\tilde{G}$, which is higher (lower) than $G$ at high redshift for positive (negative) values of the coupling constant $\xi$, while it is equal to $G$ at $z=0$ in order to recover General Relativity (see Fig. \ref{dG_z}). This means that in EQp gravity becomes weaker at low redshift compared to high redshift, while in EQn it becomes stronger. Thus one can expect that in EQp (EQn) the halos expand (contract) due to the change in the gravitational interaction, resulting in lower (higher) concentrations with respect to the case in which gravity is constant.

\noindent For the relaxed sample, compared to $\Lambda$CDM, the differences in the normalization are less than $10\%$, while the differences in the slope can almost reach $15\%$, but they are compatible with the standard errors. Also in this case, the most extreme cosmologies are EQp and EQn, whose normalizations go in the opposite direction with respect to their $\sigma_{8} D_{+}$. This fact confirms the conclusions we have drawn from the complete sample. The values of the reduced chi-squared indicate that equation (\ref{c-M}) is a good parametrization of the $c-M$ relation for almost all cosmological models. Only SUGRA has high values both for the complete and relaxed samples.

\begin{table}
\caption{Best-fit parameters, standard errors and reduced chi-squared $\tilde{\chi}^2$ of the $c-M$ relation equation (\ref{c-M}) for dark matter only density profile fit in the region [$0.1-1$]$R_{200}$ for the complete and relaxed samples of the five different cosmological models at $z=0$.}
\begin{tabular}{|cc|cc|cc|c|}
\hline Model & $\sigma_8$ & A & $\sigma_{A}$ & $B$ & $\sigma_{B}$ & $\tilde{\chi}^2$ \\
\hline & & \multicolumn{5}{|c|}{dm (all)} \\
\hline
$\Lambda$CDM & $0.776$ & $3.59$ & $0.05$ & $-0.099$ & $0.011$ & $0.48$ \\
RP & $0.746$ & $3.54$ & $0.05$ & $-0.103$ & $0.011$ & $1.14$ \\
SUGRA & $0.686$ & $3.41$ & $0.05$ & $-0.098$ & $0.013$ & $1.50$ \\
EQp & $0.748$ & $3.36$ & $0.05$ & $-0.097$ & $0.012$ & $0.35$ \\
EQn & $0.726$ & $3.70$ & $0.05$ & $-0.069$ & $0.013$ & $0.78$ \\
\hline & & \multicolumn{5}{|c|}{dm (relaxed)} \\
\hline 
$\Lambda$CDM & $0.776$ & $4.09$ & $0.05$ & $-0.092$ & $0.011$ & $0.66$ \\
RP & $0.746$ & $4.08$ & $0.05$ & $-0.081$ & $0.011$ & $0.92$ \\
SUGRA & $0.686$ & $3.94$ & $0.06$ & $-0.081$ & $0.012$ & $1.55$ \\
EQp & $0.748$ & $3.84$ & $0.05$ & $-0.097$ & $0.011$ & $1.32$ \\
EQn & $0.726$ & $4.25$ & $0.06$ & $-0.081$ & $0.013$ & $0.51$ \\
\hline
\end{tabular}
\label{tab_parameters_core_cut_dm}
\end{table}

\noindent Our results are in good qualitative agreement with the findings of \cite{2011ApJ...728..109L}, where halos in extended quintessence models have lower (higher) concentrations with respect to the $\Lambda$CDM case for positive (negative) values of the scalar field coupling.

\begin{figure*}
\hbox{
 \epsfig{figure=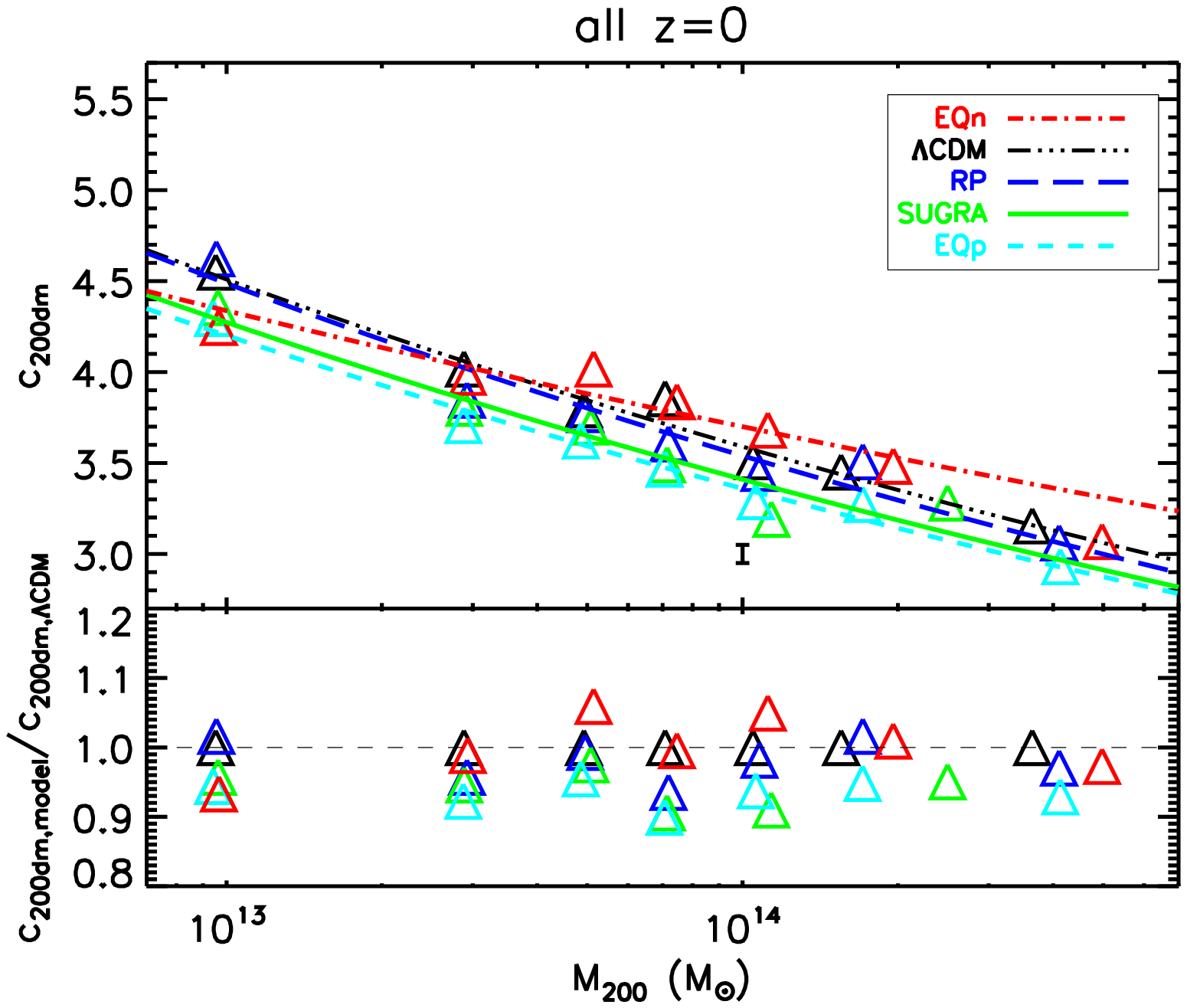,width=0.50\textwidth}
 \epsfig{figure=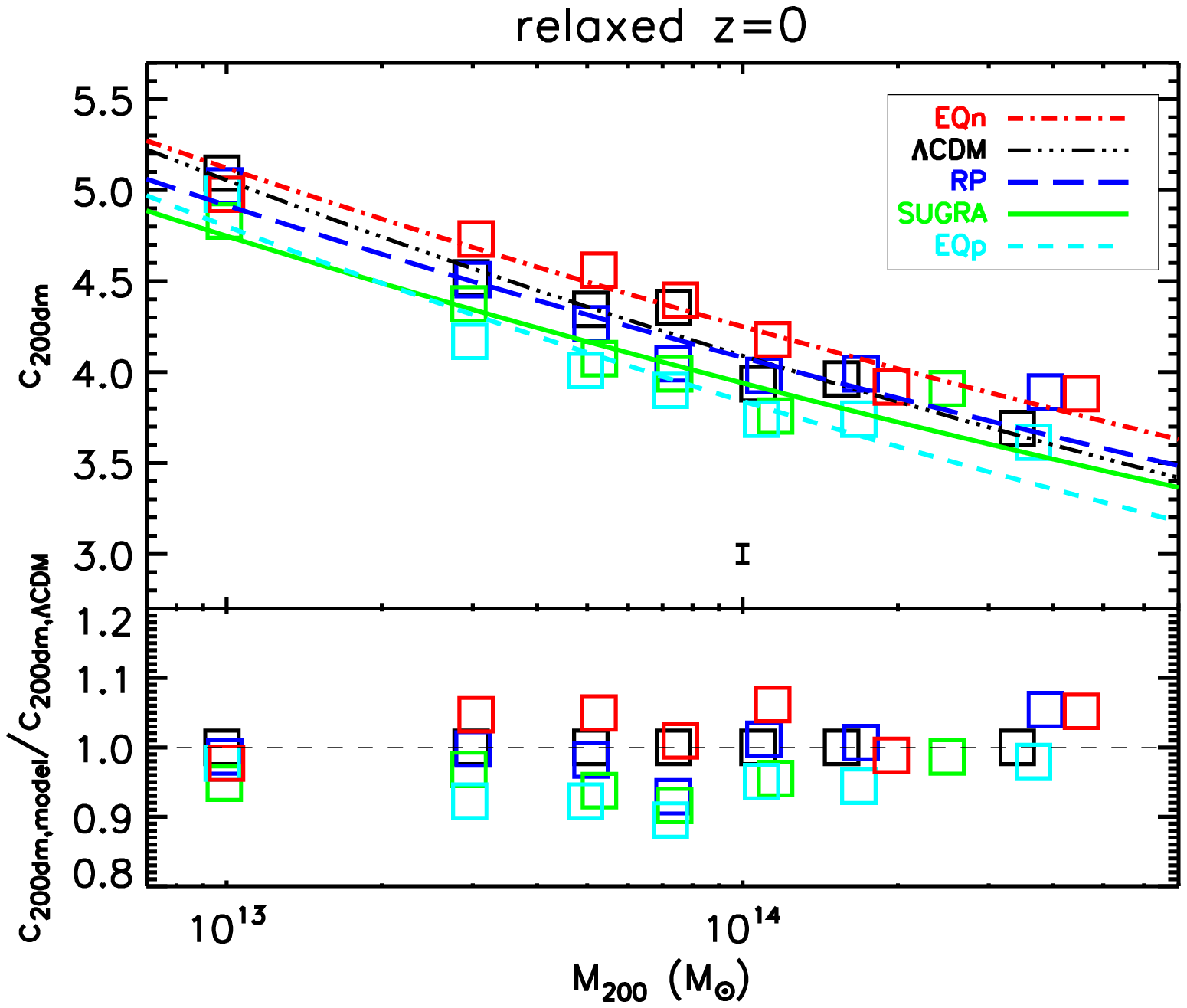,width=0.50\textwidth}
}
\hbox{ 
}
\caption{(Left panel) The values of $c_{200dm}$ for the complete sample of the $\Lambda$CDM (black), RP (blue), SUGRA (green), EQp (cyan), and EQn (red) cosmologies at $z=0$. The lines of the corresponding colours are our best-fit of $c-M$ relation equation (\ref{c-M}) and the vertical black bar is the error on the normalization of $\Lambda$CDM as listed in Table \ref{tab_comparison}. The symbols in the low part of the panel are the ratios between $c_{200dm}$ for the model and $c_{200dm}$ for $\Lambda$CDM. (Right panel) The same as in the left panel, but for the relaxed sample.}
\label{cdm-M_fit}
\end{figure*}

We plot the best-fits $c-M$ relations, along with the binned data for all the cosmological models, in Fig. \ref{cdm-M_fit}. We clearly see that the results on the normalizations are due to differences in the concentrations over a wide mass range. If we look, for example, at the complete sample (left panel of Fig. \ref{cdm-M_fit}), we see that the different slope of EQn is mainly originated by the less massive bin. But with the exception of this bin, EQn shows the highest concentration in almost all the mass bins, while in general EQp has the lowest concentration. For the relaxed sample (right panel of Fig. \ref{cdm-M_fit}), the relative behaviour of the different cosmological models is even clearer, and indeed the differences in the slopes are less pronounced.

We also try to limit the fit of the $c-M$ relation to the halos with $M_{200m} > 10^{14} \ {\rm{M_{\odot}}} \ h^{-1}$, in order to check the effect of including low-mass objects. We report the results in Table \ref{tab_parameters_mass_cut_dm}. For the complete sample, we note that the slopes and the normalizations are compatible to the fit including also low-mass objects, but that the standard errors are a factor of two (four) higher for the normalization (slope), meaning that the relation is less tight. Notable exceptions are EQn, which shows a steep slope, and SUGRA, which shows a positive trend of $c_{200dm}$ with mass (even if with large uncertainties) and a consequently low normalization. The slope of EQn can be explained by the fact that excluding from the fit the lowest mass bin, which is the one with the lowest concentration, results in a steeper slope. The behaviour of SUGRA can be explained by the fact that this model lacks very massive objects (see Paper I), that are the ones with the lower concentration. 

For the same reason, the $c-M$ relation for relaxed objects is flatter than when we include low-mass objects. This is expected, because we do not have the low-mass objects that have high concentration and we do not have the high-mass objects, which are in general more disturbed, that have low concentration. Thus, in this mass range, the relation is in general almost flat, with big uncertainties on the slope.

\begin{table}
\caption{Best-fit parameters and standard errors of the $c-M$ relation equation (\ref{c-M}) for dark matter only density profile fit in the region [$0.1-1$]$R_{200}$ for objects with $M_{200m} > 10^{14} \ {\rm{M_{\odot}}} \ h^{-1}$ in the complete and relaxed samples of the five different cosmological models at $z=0$.}
\begin{tabular}{|cc|cc|cc|}
\hline Model & $\sigma_8$ & $A$ & $\sigma_{A}$ & $B$ & $\sigma_{B}$ \\
\hline & & \multicolumn{4}{|c|}{dm (all)} \\
\hline
$\Lambda$CDM & $0.776$ & $3.55$ & $0.09$ & $-0.087$ & $0.038$ \\
RP & $0.746$ & $3.54$ & $0.11$ & $-0.080$ & $0.046$ \\
SUGRA & $0.686$ & $3.18$ & $0.12$ & $+0.035$ & $0.060$ \\
EQp & $0.748$ & $3.35$ & $0.10$ & $-0.080$ & $0.044$ \\
EQn & $0.726$ & $3.74$ & $0.11$ & $-0.114$ & $0.048$ \\
\hline & & \multicolumn{4}{|c|}{dm (relaxed)} \\
\hline 
$\Lambda$CDM & $0.776$ & $3.99$ & $0.10$ & $-0.055$ & $0.042$ \\
RP & $0.746$ & $4.00$ & $0.12$ & $-0.017$ & $0.045$ \\
SUGRA & $0.686$ & $3.73$ & $0.13$ & $+0.051$ & $0.059$ \\
EQp & $0.748$ & $3.76$ & $0.11$ & $-0.027$ & $0.046$ \\
EQn & $0.726$ & $4.19$ & $0.13$ & $-0.079$ & $0.053$ \\
\hline
\end{tabular}
\label{tab_parameters_mass_cut_dm}
\end{table}

\section{Dark energy models: results on the total profiles} \label{total}

In this section, we study the impact of baryon physics on the $c-M$ relation by analysing the hydrodynamical runs of our simulations for all the cosmological models under investigation. This allows us to understand the effects of the presence of a dynamical dark energy component on the internal matter distribution, including baryons, of the halos. We start this analysis with the objects at $z=0$, then we will consider the redshift evolution of the $c-M$ relation by including also objects at $z=0.5$ and $z=1$. As we already explained in Sect. \ref{fitting}, first of all we fit equation (\ref{c-M}) for the objects at $z=0$, then, keeping fixed the best-fit values of $A$ and $B$, we fit equation (\ref{c-M_z}) and evaluate the redshift evolution $C$.

In Table \ref{tab_parameters_core_cut_tot} we summarize the best-fit parameters, the standard errors and the reduced chi-squared of the $c-M$ relations equation (\ref{c-M}) and equation (\ref{c-M_z}) for the five cosmological models here considered, both for the complete and relaxed samples. For all the cosmologies, the values of $A$ are larger than in the dark matter only case, indicating that the inclusion of baryons leads to an increase of the concentration, while the standard errors remain the same. The slope is somewhat flatter than in the dark matter only case for all the cosmological models. We already noted both these features in the $\Lambda$CDM case (see Sect. \ref{fitting}): the shallower slope can be explained by the fact that the increase in the concentration due to the presence of baryons is greater in high-mass objects than in low-mass ones. For the complete sample, the total $c-M$ relation reflects the one for dark matter, with the normalizations in the same order, with the exception of an exchange between EQp and SUGRA. The relaxed sample shows higher normalizations than the complete sample, as in the dark matter only case, while the slopes are very similar to the ones of the complete sample. Also in this case, in comparison with the order of $\sigma_{8} D_{+}$, the extreme case are EQp and EQn. Thus, the trend we find in the dark matter only runs still holds in the hydrodynamical runs. The values of the reduced chi-squared indicate that for some models equation (\ref{c-M}) is a good parametrization also for the $c-M$ relation of objects extracted from the hydrodynamical runs, at least when the complete sample is considered. For relaxed objects only SUGRA seems to be described quite well by this relation.

\begin{table*}
\caption{Best-fit parameters, standard errors and reduced chi-squared $\tilde{\chi}^2$ of the $c-M$ relation equation (\ref{c-M}) for total density profile fit in the region [$0.1-1$]$R_{200}$ for the complete and relaxed samples of the five different cosmological models at $z=0$, and of the $c-M$ relation equation (\ref{c-M_z}) including also objects at $z=0.5$ and $z=1$, keeping the best-fit values of $A$ and $B$ found at $z=0$ fixed.}
\begin{tabular}{|cc|cc|cc|c|cc|c|}
\hline Model & $\sigma_8$ & $A$ & $\sigma_{A}$ & $B$ & $\sigma_{B}$ & $\tilde{\chi}^2$ & $C$ & $\sigma_{C}$ & $\tilde{\chi}^2$ \\
\hline & & \multicolumn{8}{|c|}{total (all)} \\
\hline
$\Lambda$CDM & $0.776$ & $3.81$ & $0.05$ & $-0.061$ & $0.011$ & $1.69$ & $-0.26$ & $0.02$ & $2.40$ \\
RP & $0.746$ & $3.72$ & $0.05$ & $-0.073$ & $0.012$ & $1.06$ & $-0.15$ & $0.02$ & $1.98$ \\
SUGRA & $0.686$ & $3.68$ & $0.06$ & $-0.057$ & $0.013$ & $0.70$ & $-0.05$ & $0.02$ & $3.68$ \\
EQp & $0.748$ & $3.69$ & $0.05$ & $-0.085$ & $0.012$ & $0.88$ & $-0.20$ & $0.02$ & $2.99$ \\
EQn & $0.726$ & $3.94$ & $0.05$ & $-0.052$ & $0.012$ & $1.55$ & $-0.21$ & $0.02$ & $1.77$ \\
\hline & & \multicolumn{8}{|c|}{total (relaxed)} \\
\hline
$\Lambda$CDM & $0.776$ & $4.29$ & $0.05$ & $-0.064$ & $0.011$ & $1.54$ & $-0.26$ & $0.02$ & $1.97$ \\
RP & $0.746$ & $4.24$ & $0.05$ & $-0.075$ & $0.010$ & $2.86$ & $-0.15$ & $0.02$ & $1.84$ \\
SUGRA & $0.686$ & $4.25$ & $0.06$ & $-0.045$ & $0.011$ & $1.14$ & $-0.08$ & $0.02$ & $1.11$ \\
EQp & $0.748$ & $4.13$ & $0.05$ & $-0.091$ & $0.010$ & $2.02$ & $-0.16$ & $0.02$ & $2.79$ \\
EQn & $0.726$ & $4.48$ & $0.05$ & $-0.057$ & $0.012$ & $1.62$ & $-0.16$ & $0.02$ & $1.02$ \\
\hline
\end{tabular}
\label{tab_parameters_core_cut_tot}
\end{table*}

\begin{figure*}
\hbox{
 \epsfig{figure=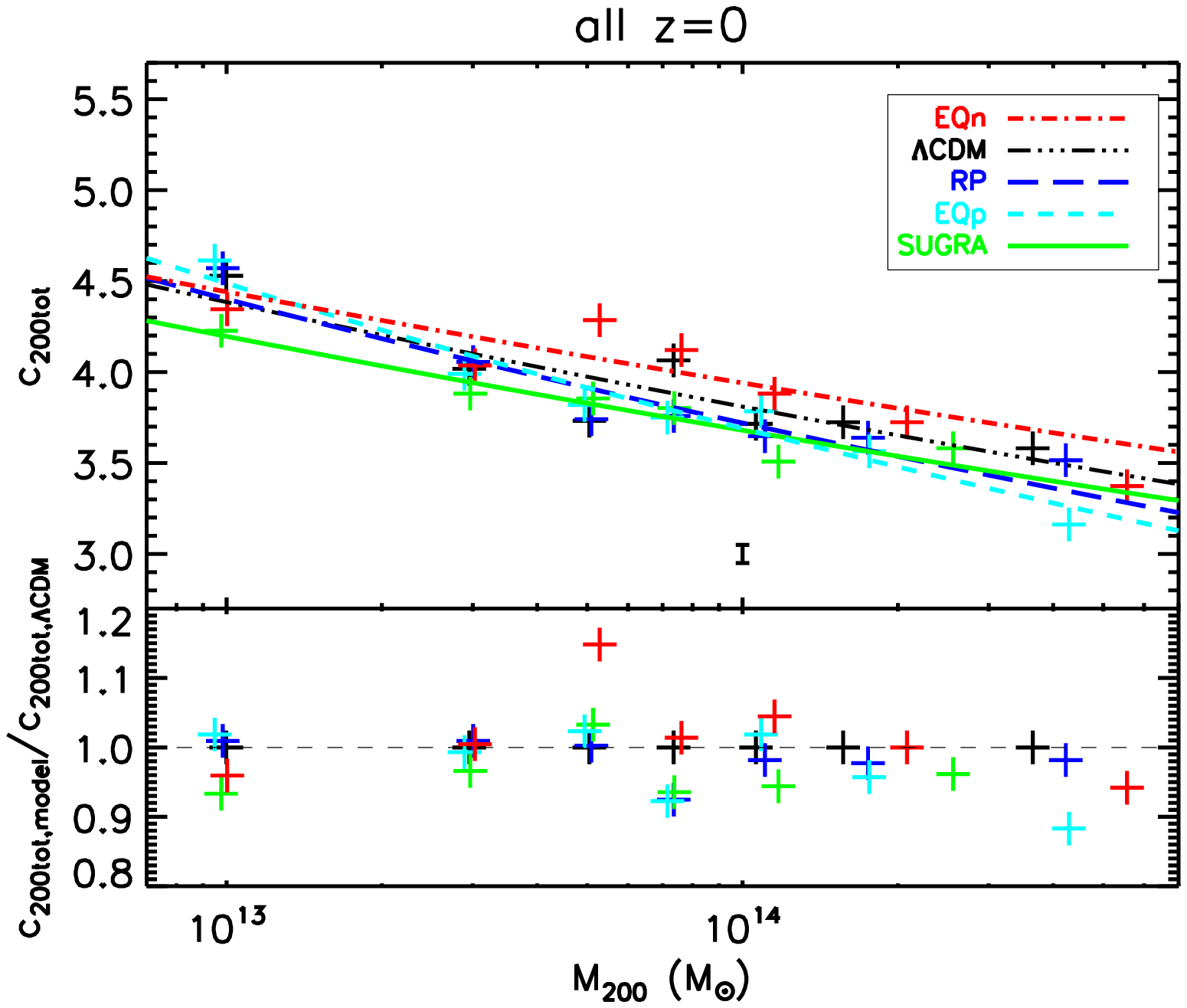,width=0.50\textwidth}
 \epsfig{figure=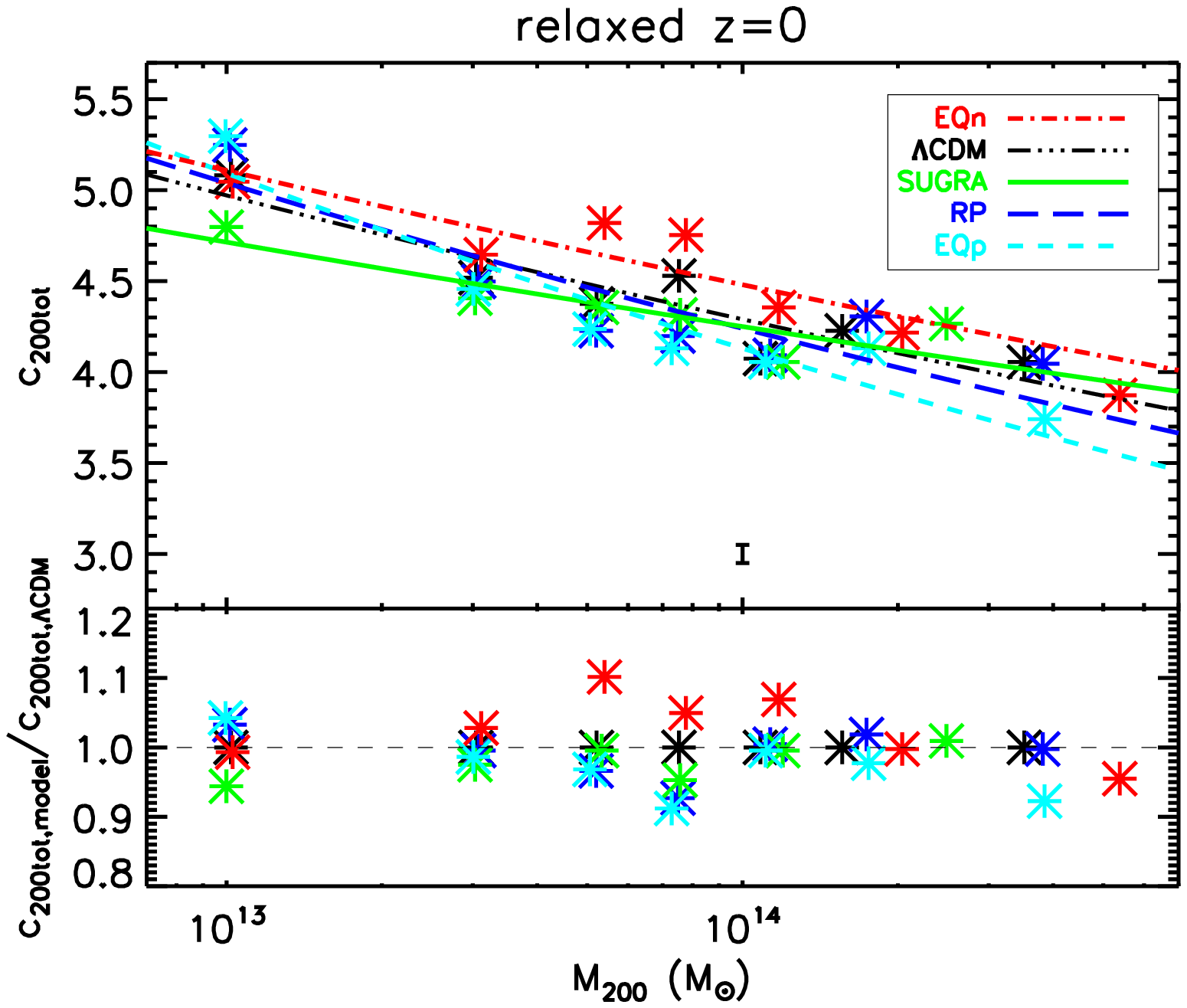,width=0.50\textwidth}
}
\hbox{ 
}
\caption{(Left panel) The values of $c_{200tot}$ for the complete sample of the $\Lambda$CDM (black), RP (blue), SUGRA (green), EQp (cyan), and EQn (red) cosmologies at $z=0$. The lines of the corresponding colours are our best-fit of $c-M$ relation equation (\ref{c-M}) and the vertical black bar is the error on the normalization of $\Lambda$CDM as listed in Table \ref{tab_comparison}. The symbols in the low part of the panel are the ratios between $c_{200tot}$ for the model and $c_{200tot}$ for $\Lambda$CDM. (Right panel) The same as in the left panel, but for the relaxed sample.}
\label{ctot-M_fit}
\end{figure*}

We plot the best-fit $c-M$ relations, along with the binned data for all the cosmological models, in Fig. \ref{ctot-M_fit}. Also in this case, we see that differences in the concentration are present over a large mass range. For the complete sample, the lower normalization of SUGRA compared to EQp is mainly due to low-mass objects, which flatten the relation. The same flattening happens for the relaxed sample, leading in this case to a higher normalization for SUGRA with respect to RP.

As for the dark matter only case, we also limit our analysis at $z=0$ also to objects with $M_{200m} > 10^{14} \ {\rm{M_{\odot}}} \ h^{-1}$. We report the results in Table \ref{tab_parameters_mass_cut_tot}. We find that the normalizations are similar to the one found including also low-mass objects, but the standard errors are higher, and the slopes are generally flatter, with some exceptions. The most evident is SUGRA, for which the trend of $c_{200}$ with mass is increasing both in the complete and relaxed samples. For the complete sample, EQp and EQn show a steepening in the slope, a behaviour which is maintained by EQn also when considering the relaxed sample. For SUGRA, the explanation is the same as in the dark matter only case, {\it{i.e.}} the lack of very high mass, low-concentration systems. For EQp and EQn, they show objects with low concentration in particular in the low-mass tail, so when excluding these objects, the result is a steepening of the slope (see Fig. \ref{ctot-M_fit}). In all cases, the scatter in the relation is considerably higher than when we consider also low-mass objects, up to a factor of three (four) in normalization (slope).

\begin{table}
\caption{Best-fit parameters and standard errors of the $c-M$ relation equation (\ref{c-M}) for total density profile fit in the region [$0.1-1$]$R_{200}$ for objects with $M_{200m} > 10^{14} \ {\rm{M_{\odot}}} \ h^{-1}$ in the complete and relaxed samples of the five different cosmological models at $z=0$.}
\begin{tabular}{|cc|cc|cc|}
\hline Model & $\sigma_8$ & $A$ & $\sigma_{A}$ & $B$ & $\sigma_{B}$ \\
\hline & & \multicolumn{4}{|c|}{total (all)} \\
\hline
$\Lambda$CDM & $0.776$ & $3.74$ & $0.09$ & $-0.031$ & $0.034$ \\
RP & $0.746$ & $3.67$ & $0.11$ & $-0.027$ & $0.040$ \\
SUGRA & $0.686$ & $3.49$ & $0.13$ & $+0.027$ & $0.061$ \\
EQp & $0.748$ & $3.83$ & $0.10$ & $-0.130$ & $0.041$ \\
EQn & $0.726$ & $3.94$ & $0.12$ & $-0.083$ & $0.045$ \\
\hline & & \multicolumn{4}{|c|}{total (relaxed)} \\
\hline 
$\Lambda$CDM & $0.776$ & $4.14$ & $0.10$ & $-0.007$ & $0.036$ \\
RP & $0.746$ & $4.17$ & $0.11$ & $-0.004$ & $0.041$ \\
SUGRA & $0.686$ & $4.00$ & $0.15$ & $+0.069$ & $0.061$ \\
EQp & $0.748$ & $4.14$ & $0.11$ & $-0.054$ & $0.043$ \\
EQn & $0.726$ & $4.41$ & $0.14$ & $-0.068$ & $0.052$ \\
\hline
\end{tabular}
\label{tab_parameters_mass_cut_tot}
\end{table}

For the hydrodynamical simulations, we also study the redshift evolution of the $c-M$ relation by fitting equation (\ref{c-M_z}) keeping the values of $A$ and $B$ fixed to the best-fit values at $z=0$. We report the results in Table \ref{tab_parameters_core_cut_tot}. For all the models, the evolution with redshift is negative, meaning that objects at higher redshifts have also lower concentrations. For the complete sample, the redshift dependence is stronger for $\Lambda$CDM than for the other cosmological models. The two EQ models have similar dependences, lower than $\Lambda$CDM, followed by RP and SUGRA, which show a very weak evolution with redshift. If we focus on the relaxed sample, we see that the $\Lambda$CDM model still shows the strongest redshift dependence, analogous to the one for the complete sample. RP, EQp and EQn have similar evolutions, but while for RP it is the same as for the complete sample, for EQ models it is weaker. SUGRA has a flatter dependence compared to the other models, but more pronounced than for the complete sample. The values of the reduced chi-squared are rather high for all the cosmological models, both for the complete and relaxed samples. Two notable exceptions are the values for the relaxed samples of SUGRA and EQn. 
\noindent The results shown in Table \ref{tab_parameters_core_cut_tot} seem to indicate that equation (\ref{c-M}), and its generalization equation (\ref{c-M_z}), are only a limited parametrization of the $c-M$ relation of galaxy clusters extracted from hydrodynamical runs of cosmological simulations including dynamical dark energy, and of its redshift evolution.

\begin{figure}
\hbox{
 \epsfig{figure=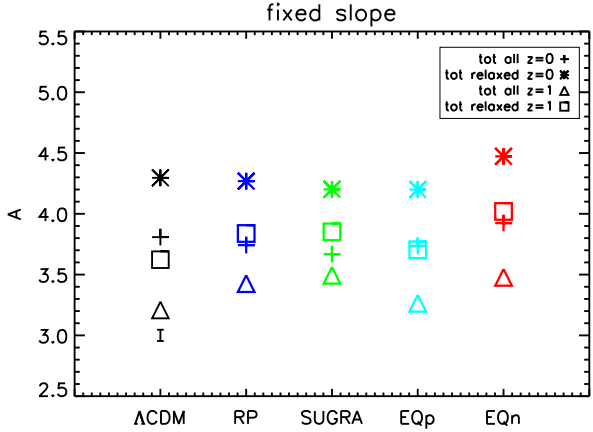,width=0.50\textwidth}
}
\caption{Best-fit normalization comparison for equation (\ref{c-M}) with $B$ fixed at the best-fit value for the complete sample of $\Lambda$CDM at $z=0$ for the $\Lambda$CDM (black), RP (blue), SUGRA (green), EQp (cyan), and EQn (red) cosmologies. Crosses: total profile fit $z=0$, complete sample. Stars: total profile fit $z=0$, relaxed sample. Triangles: total profile fit $z=1$, complete sample. Squares: total profile fit $z=1$, relaxed sample. The vertical black bar is the error on the normalization of the complete sample of $\Lambda$CDM at $z=0$.}
\label{normalization_z1}
\end{figure}

Finally, we check the evolution with redshift of the normalization $A$ both for the complete and relaxed samples for all the cosmological models. We do that by fitting the $c-M$ relation equation (\ref{c-M}) by keeping fixed the slope at the best-fit value for the complete sample of $\Lambda$CDM at $z=0$ ({\it{i.e.}} $B=-0.061$, see Table \ref{tab_parameters_core_cut_tot}) and leaving only $A$ as a free parameter. We compare the different normalizations found by fitting equation (\ref{c-M}) in this way for both the complete and relaxed samples at $z=0$ and at $z=1$. Thus we can have a snapshot of the imprint on dark energy on the concentration of the halos. We summarize the results in Table \ref{tab_normalization_z1} and plot them in Fig. \ref{normalization_z1}. Indeed we see that, for a given sample at $z=0$, the normalization is decreasing going from $\Lambda$CDM to RP to SUGRA, as expected from $\sigma_{8} D_{+}$. Then, as we already discussed, the normalization of EQp is rather suppressed with respect to this simple expectation, while the one of EQn is enhanced, due to the evolution of the effective gravitational interaction $\tilde{G}$ in these models. If we move instead to $z=1$, the relative behaviour of the different cosmological models changes. We find that, for both samples, the normalization is increasing going from $\Lambda$CDM to RP to SUGRA, while EQp is still suppressed and EQn is still enhanced. The findings on $\Lambda$CDM, RP and SUGRA at $z=1$ are in agreement with the strong redshift evolution we found for $\Lambda$CDM and with the weak redshift evolution we found for SUGRA. There is an evolution from a low-normalization to a high-normalization relation for the first model, and vice versa for the latter. We show the values of the reduced chi-squared of the fit as a reference, but we do not discuss them because we are imposing the slope for $\Lambda$CDM also to other models.

\begin{table*}
\caption{Best-fit parameters, standard errors and reduced chi-squared $\tilde{\chi}^2$ of the $c-M$ relation equation (\ref{c-M}), with $B$ fixed at the best-fit value for the complete sample of $\Lambda$CDM at $z=0$, for total density profile fit in the region [$0.1-1$]$R_{200}$ for the complete and relaxed samples of the five different cosmological models at $z=0$ and $z=1$.}
\begin{tabular}{|cc|ccc|ccc|ccc|ccc|}
\hline & & \multicolumn{6}{|c|}{$z=0$} & \multicolumn{6}{|c|}{$z=1$} \\
\hline Model & $\sigma_8$ & $A$ & $\sigma_{A}$ & $\tilde{\chi}^2$ & $A$ & $\sigma_{A}$ & $\tilde{\chi}^2$ & $A$ & $\sigma_{A}$ & $\tilde{\chi}^2$ & $A$ & $\sigma_{A}$ & $\tilde{\chi}^2$ \\
\hline & & \multicolumn{3}{|c|}{total (all)} & \multicolumn{3}{|c|}{total (relaxed)} & \multicolumn{3}{|c|}{total (all)} & \multicolumn{3}{|c|}{total (relaxed)} \\
\hline
$\Lambda$CDM & $0.776$ & $3.81$ & $0.05$ & $1.69$ & $4.30$ & $0.05$ & $1.30$ & $3.21$ & $0.05$ & $6.43$ & $3.62$ & $0.06$ & $2.09$ \\
RP & $0.746$ & $3.74$ & $0.05$ & $1.08$ & $4.27$ & $0.05$ & $2.69$ & $3.43$ & $0.06$ & $4.27$ & $3.84$ & $0.08$ & $2.05$ \\
SUGRA & $0.686$ & $3.67$ & $0.05$ & $0.58$ & $4.20$ & $0.05$ & $1.28$ & $3.49$ & $0.07$ & $1.81$ & $3.85$ & $0.08$ & $0.43$ \\
EQp & $0.748$ & $3.74$ & $0.04$ & $1.49$ & $4.20$ & $0.05$ & $3.13$ & $3.26$ & $0.06$ & $6.09$ & $3.70$ & $0.07$ & $2.72$ \\
EQn & $0.726$ & $3.92$ & $0.05$ & $1.38$ & $4.47$ & $0.05$ & $1.37$ & $3.48$ & $0.07$ & $3.43$ & $4.02$ & $0.08$ & $1.69$ \\
\hline
\end{tabular}
\label{tab_normalization_z1}
\end{table*}

\section{Discussion} \label{other}

\begin{figure}
\hbox{
 \epsfig{figure=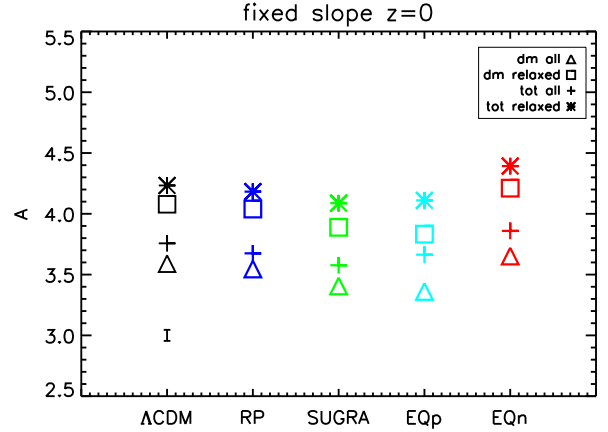,width=0.50\textwidth}
}
\caption{Best-fit normalization comparison for equation (\ref{c-M}) with $B$ fixed at the best-fit value for the dark matter only complete sample of $\Lambda$CDM at $z=0$ for the $\Lambda$CDM (black), RP (blue), SUGRA (green), EQp (cyan), and EQn (red) cosmologies. Triangles: dark matter profile fit, complete sample. Squares: dark matter profile fit, relaxed sample. Crosses: total profile fit, complete sample. Stars: total profile fit, relaxed sample. The vertical black bar is the error on the normalization of the dark matter only complete sample of $\Lambda$CDM.}
\label{normalization}
\end{figure}

In this section, we discuss the results we have found for the dark matter only and hydrodynamical runs for the different cosmologies. We start by comparing the normalization of the $c-M$ relation at $z=0$ for the complete and relaxed samples in both runs. To do this, we fix the slope at the best-fit value for the complete sample of the dark matter only run for $\Lambda$CDM at $z=0$ ({\it{i.e.}} $B=-0.099$, see Table \ref{tab_parameters_core_cut_dm}) and we fit equation (\ref{c-M}) with only $A$ as a free parameter. We report the results in Table \ref{tab_normalization} and plot them in Fig. \ref{normalization}, which summarizes almost all the information on the $c-M$ relation at $z=0$ for the cosmological models under investigation. Indeed, we see that inside each sample, objects in the dark matter only runs have a lower normalization than objects in the hydrodynamical runs, independently of cosmology and dynamical state. Moreover, inside each run, relaxed objects have a higher normalization compared to the complete sample. Then, as a general trend, the normalization is decreasing going from $\Lambda$CDM to RP to SUGRA, independently of the run and the dynamical state. Finally EQn always has the highest normalization while EQp, at least for the dark matter only runs, alway has the lowest. In particular, EQn has the highest normalization also in hydrodynamical runs, while the suppression of the concentration in EQp is somehow mitigated in these runs. The behaviour of $\Lambda$CDM, RP and SUGRA is in agreement with the simple idea that the normalization of the $c-M$ relation is driven by the value of $\sigma_{8} D_{+}$ (but the one of EQp and EQn is not). We ascribe the behaviour of the two EQ models to the different redshift evolution of the effective gravitational interaction $\tilde{G}$, which decreases, going from high to low redshift, toward the General Relativity value for EQp, while it increases for EQn (see Sect. \ref{models}). This leads to a lower concentration, and thus to a lower normalization, for EQn halos and to a higher concentration, and thus a higher normalization, for EQp halos (see Sect. \ref{dark_matter}). We show the values of the reduced chi-squared of the fit as a reference, but we do not discuss them because we are imposing the slope for $\Lambda$CDM also to other models.

\begin{table*}
\caption{Best-fit parameters, standard errors and reduced chi-squared $\tilde{\chi}^2$ of the $c-M$ relation equation (\ref{c-M}), with $B$ fixed at the best-fit value for the dark matter only complete sample of $\Lambda$CDM at $z=0$, for dark matter only and total density profile fit in the region [$0.1-1$]$R_{200}$ for the complete and relaxed samples of the five different cosmological models at $z=0$.}
\begin{tabular}{|cc|ccc|ccc|ccc|ccc|}
\hline Model & $\sigma_8$ & $A$ & $\sigma_{A}$ & $\tilde{\chi}^2$ & $A$ & $\sigma_{A}$ & $\tilde{\chi}^2$ & $A$ & $\sigma_{A}$ & $\tilde{\chi}^2$ & $A$ & $\sigma_{A}$ & $\tilde{\chi}^2$ \\
\hline & & \multicolumn{3}{|c|}{dm (all)} & \multicolumn{3}{|c|}{dm (relaxed)} & \multicolumn{3}{|c|}{total (all)} & \multicolumn{3}{|c|}{total (relaxed)} \\
\hline
$\Lambda$CDM & $0.776$ & $3.59$ & $0.05$ & $0.48$ & $4.08$ & $0.04$ & $0.62$ & $3.76$ & $0.04$ & $3.36$ & $4.23$ & $0.05$ & $3.06$ \\
RP & $0.746$ & $3.55$ & $0.04$ & $0.97$ & $4.04$ & $0.05$ & $1.19$ & $3.67$ & $0.04$ & $1.66$ & $4.18$ & $0.05$ & $3.29$ \\
SUGRA & $0.686$ & $3.41$ & $0.04$ & $1.20$ & $3.89$ & $0.04$ & $1.69$ & $3.58$ & $0.04$ & $2.77$ & $4.09$ & $0.04$ & $5.56$ \\
EQp & $0.748$ & $3.36$ & $0.04$ & $0.30$ & $3.83$ & $0.04$ & $1.11$ & $3.66$ & $0.04$ & $0.96$ & $4.11$ & $0.04$ & $1.77$ \\
EQn & $0.726$ & $3.65$ & $0.04$ & $1.56$ & $4.21$ & $0.05$ & $0.74$ & $3.86$ & $0.05$ & $3.81$ & $4.39$ & $0.05$ & $3.46$ \\
\hline
\end{tabular}
\label{tab_normalization}
\end{table*}

\begin{figure*}
\hbox{
 \epsfig{figure=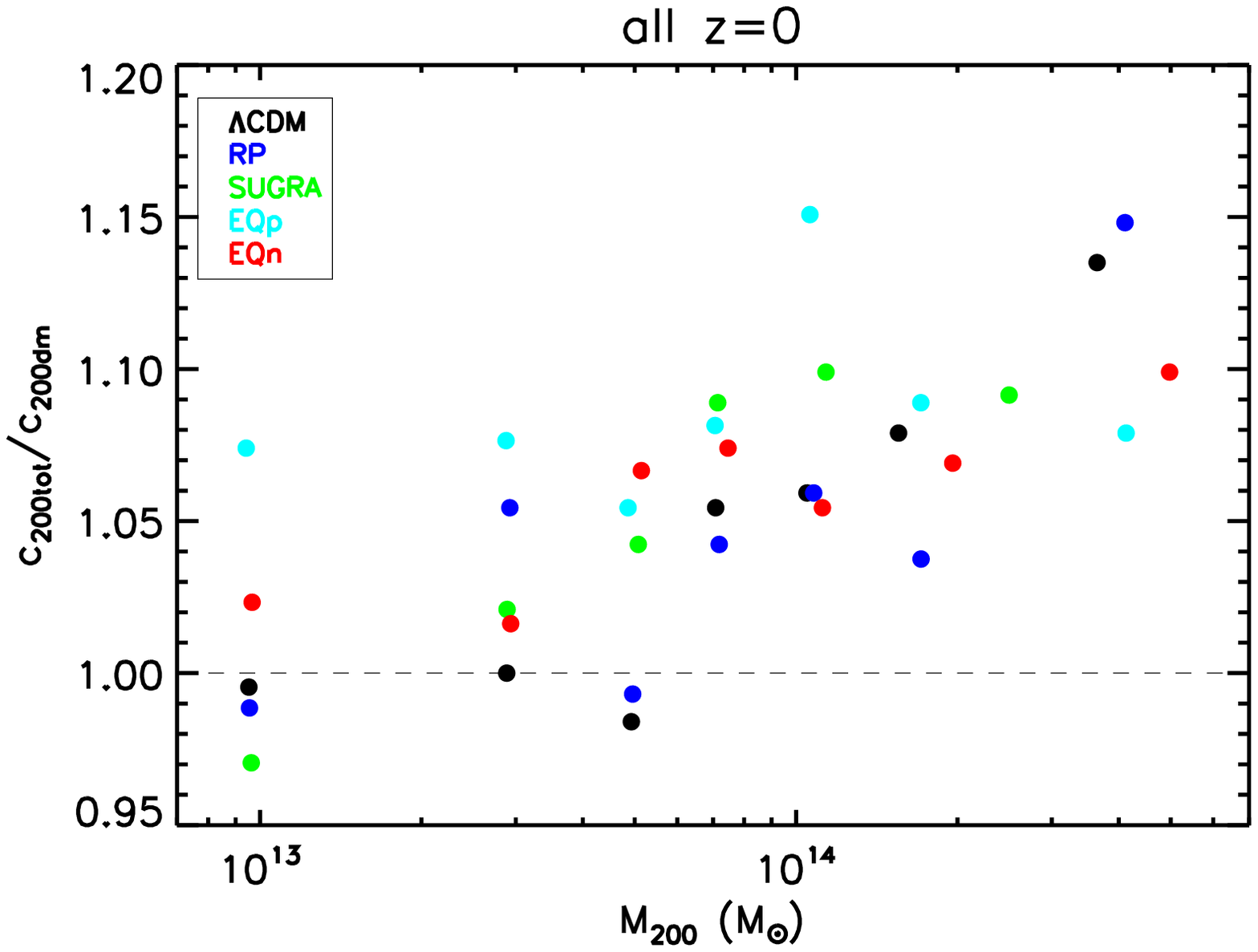,width=0.50\textwidth}
 \epsfig{figure=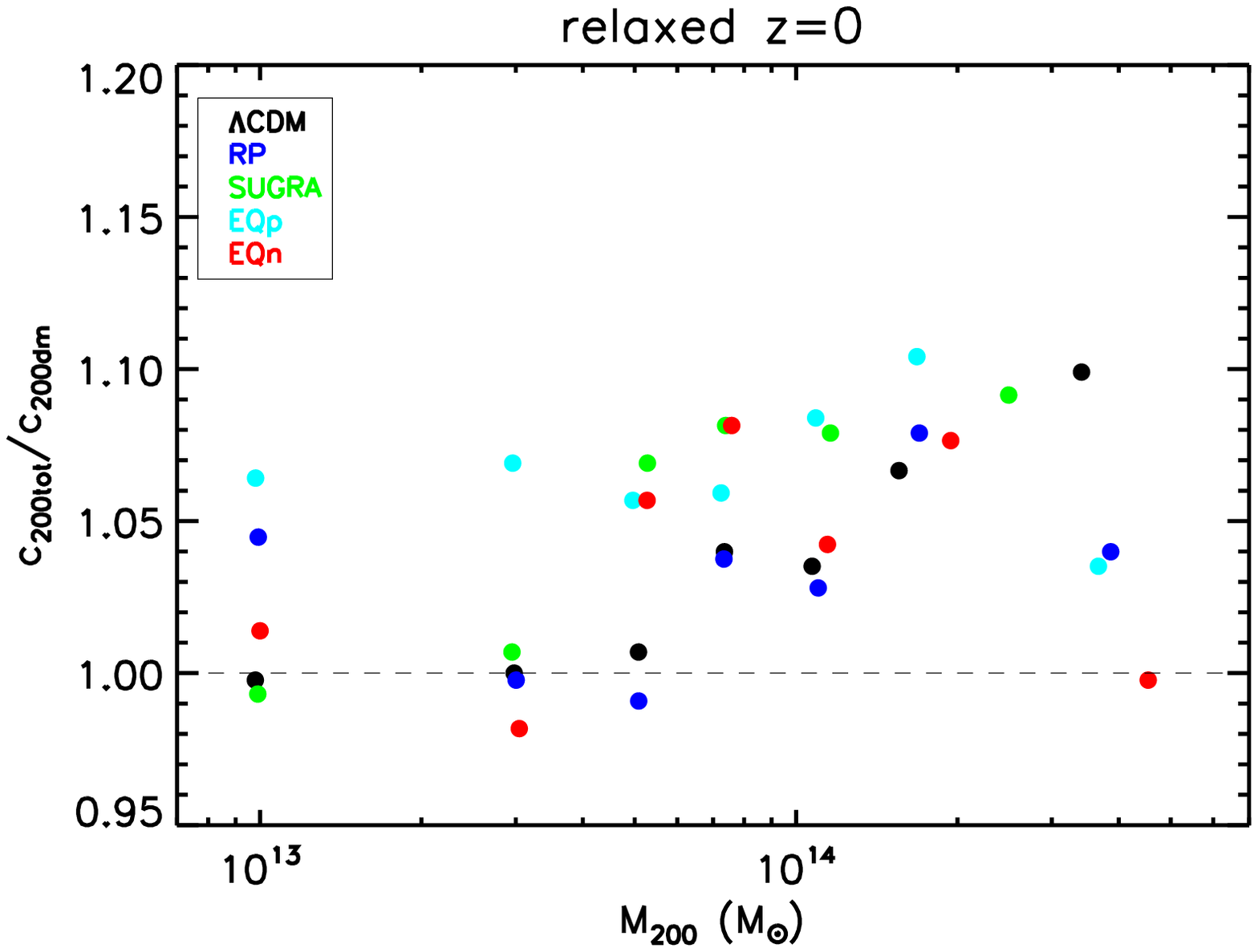,width=0.50\textwidth}
}
\caption{(Left panel) Ratio between $c_{200tot}$ and $c_{200dm}$ for the complete sample of the $\Lambda$CDM (black), RP (blue), SUGRA (green), EQp (cyan), and EQn (red) cosmologies at $z=0$. (Right panel) The same as in the left panel, but for the relaxed sample.}
\label{cdm_vs_ctot}
\end{figure*}

Finally, we compare the concentrations obtained from the dark matter only runs with the one obtained from the hydrodynamical runs. We do this by taking the ratio between $c_{200tot}$ and $c_{200dm}$ in each mass bin both for the complete and relaxed samples at $z=0$. We plot the results in Fig. \ref{cdm_vs_ctot}. In the left panel we analyse the ratio for the complete sample. We see that all cosmological models have $c_{200tot}/c_{200dm}$ increasing with increasing mass. This fact confirms that the baryon physics, as treated in these numerical runs, influences more the concentration of high-mass objects compared to low-mass ones (see also Fig. \ref{dm_vs_tot_comparison}). In particular, while $\Lambda$CDM, RP and SUGRA have $c_{200tot} < c_{200dm}$ in some low-mass bins, EQp and EQn have in general $c_{200tot} > c_{200dm}$, with a less pronounced evolution with mass. In the right panel we analyse the ratio for the relaxed sample. Here the situation is a bit different. All the cosmological models still show a general increase of $c_{200tot}/c_{200dm}$ with increasing mass, but the evolution is rather shallow. In particular, very massive objects in the RP, EQp and EQn models have low values of these ratio. In any case, with few exceptions in some mass bins, we generally find that $c_{200tot} > c_{200dm}$ for all models. This analysis indicates that the inclusion of baryon physics in the simulations is unable to solve the discrepancy between the predicted and observed $c-M$ relation by increasing the slope. Indeed, in none of the cases we have analysed, the effect of the baryons is to increase the concentration of low-mass objects without affecting the one of the high-mass ones. Of course a possible explanation of this fact can be that we do not include some kinds of feedback in our simulations, in particular AGN feedback. Still, \cite{2010MNRAS.405.2161D} already showed that none of the different hydrodynamical treatments they tried was able to both explain the observed $c-M$ relation and the stellar fraction in galaxy clusters.

Nevertheless, we have seen that different dark energy models leave an imprint in the $c-M$ relation. In particular, as already noted by \cite{2004A&A...416..853D}, in ordinary quintessence models, like our RP and SUGRA, the normalization of this relation compared to the $\Lambda$CDM one is driven by the value of $\sigma_{8} D_{+}(z_{coll})$ with respect to the value for $\Lambda$CDM. We find that the same thing is no longer valid for extended quintessence models like our EQp and EQn. Indeed, in these cases we find that, in presence of a positive coupling, the value of the normalization is lower than what expected from $\sigma_{8} D_{+}(z_{coll})$ while in presence of a negative coupling it is increased. We interpret this behaviour as due to the different redshift evolution of the effective gravitational interaction $\tilde{G}$, as explained in Sect. \ref{models}. Unfortunately, given the discrepancies between the observed and the predicted $c-M$ relation, it is difficult to use the observed $c-M$ relation to disentangle different cosmological models through the imprints left by dark energy.

\section{Summary} \label{c-M_relation_summary}

In this paper we studied the $c-M$ relation for the halos extracted from the simulation set introduced in Paper I. From our analysis we draw the following conclusions.

\begin{itemize}
\item Concentration: the definition itself of the concentration of a halo can lead to very different results. For the reference $\Lambda$CDM model at $z=0$, we see that the usual fit of a NFW profile, depending on the way in which the errors are weighted and the concentration is evaluated from the fit, gives values of the concentration with differences, at worst, of the order of $5\%$. The method using $V_{max}$ discussed in \cite{2011arXiv1104.5130P}, instead, returns similar values of the concentration only for super-relaxed objects, while in general gives higher values of the concentration, up to $20\%$. The quantity $R_{2500}/R_{500}$, which is model independent, is found to be well correlated with the NFW value of the concentration for dark matter only simulations, indicating that on average dark matter halos are indeed well described by a NFW profile. In a given mass bin, the mean concentration is higher and the intrinsic scatter is lower for objects that are in a relaxed dynamical state and well described by a NFW profile. Halos extracted from the hydrodynamical runs have, in general, higher concentrations than their dark matter only counterparts. The effect is more evident in high-mass systems, due to a higher baryon fraction in the inner regions. 
\item $c-M$ relation: limiting our analysis to the $\Lambda$CDM model, there is a large intrinsic scatter in the values of the concentration for objects of a given mass, ranging from $15\%$ up to $40\%$ depending on mass and dynamical state. Nevertheless, if we consider the mean value of concentration in a given mass range, there is a good correlation between concentration and mass. The concentration is slightly decreasing with increasing mass, and this relation is well described by a power-law, with two free parameters, the normalization and the slope. The normalization, that we define as the value of the concentration of a halo with a mass of $10^{14} \ {\rm{M_{\odot}}}$, is increasing going from $A=3.59$ for the complete sample to $4.09$ and $4.52$ in the relaxed and super-relaxed samples, respectively. The slope is similar for the different samples considered and is $B \sim -0.1$. In general, we find a good agreement with the results from other works in literature. The normalization is higher by $5-15 \%$ for objects extracted from the hydrodynamical runs compared to the one of the dark matter only runs. The slope is flatter by $30 \%$ for the hydrodynamical simulations than for the dark matter only ones. This is expected because massive objects from the hydrodynamical simulations have higher concentrations than their dark matter only counterparts. 
\item Redshift dependence: the $c-M$ relation shows an evolution with redshift, with concentration decreasing with increasing redshift as $(1+z)^{-0.3}$. For the $\Lambda$CDM model, considering also objects at $z=0.5$ and $z=1$, the redshift evolution is more pronounced for the dark matter only simulations than for the hydrodynamical ones, while it is similar for the complete and relaxed samples.
\item Dark energy models: we find that the normalization of the $c-M$ relation in dynamical dark energy cosmologies is different with respect to the $\Lambda$CDM one, while the slopes are more compatible. In particular, at $z=0$, the differences in the normalizations for RP and SUGRA when compared to $\Lambda$CDM reflect the differences in $\sigma_{8}D_{+}$, with models having a higher $\sigma_{8}D_{+}$ also having a higher normalization. This simple scheme is not valid for the EQp and EQn scenarios. In the former case, the normalization is lower than expected considering $\sigma_{8} D_{+}$, while in the latter it is higher, and indeed EQn is always the model with the highest normalization, regardless of the dynamical state of the objects or the runs they are extracted from. This behaviour is due to the different redshift evolution of the effective gravitational interaction $\tilde{G}$ in these models. Indeed, going from high to low redshift, $\tilde{G}$ decreases (increases) for EQp (EQn) toward the value for $\Lambda$CDM and ordinary quintessence models, thus decreasing (increasing) the concentration of the halos. This is a very important result because it shows a direct manifestation of the coupling between gravity and the quintessence scalar field. For objects extracted from the hydrodynamical runs, we also study the redshift evolution of the $c-M$ relation. We find different evolution for different dark energy models, with values of $C$ ranging between $-0.05$ and $-0.26$. $\Lambda$CDM has the strongest evolution, while SUGRA has the weakest, while RP lies in between. For these three models, the behaviour is similar for the complete and relaxed samples. EQp and EQn models, instead, show an evolution similar to $\Lambda$CDM for the complete sample and similar to RP for the relaxed sample. It is interesting to note that at $z=0$ the normalization decreases from $\Lambda$CDM to RP to SUGRA, while at $z=1$ the situation is completely reversed, indicating a stronger redshift evolution in SUGRA with respect to $\Lambda$CDM. Independently of redshift, EQp has always the lowest normalization while EQn has always the highest, meaning that the effect of the coupling between gravity and the quintessence scalar field is important also at higher redshift.
\end{itemize}

\section*{Acknowledgments}

Computations have been performed at the ``Leibniz-Rechenzentrum''
with CPU time assigned to the Project ``h0073''.
We acknowledge financial contributions from contracts ASI I/016/07/0 COFIS,
ASI-INAF I/023/05/0, ASI-INAF I/088/06/0, ASI
`EUCLID-DUNE' I/064/08/0, PRIN MIUR 2009 ``Dark energy and cosmology with
large galaxy survey'', and PRIN INAF 2009 ``Towards an Italian network of computational cosmology''.
K.~D.~acknowledges the support of the DFG Priority Programme 1177
and additional support by the DFG Cluster of Excellence ``Origin
and Structure of the Universe''. We thank the referee Baojiu Li for helping us improving the presentation of the results of our paper. We thank Amina Helmi, Margherita Ghezzi, Mauro Roncarelli, Carlo Giocoli, Massimo Meneghetti, Stefano Borgani, Valeria Pettorino, Francesco Pace, Matthias Bartelmann, Andrea Macci\`o and Marco Baldi for useful discussions.

\end{document}